\documentclass[12pt,preprint]{aastex}

\slugcomment{To appear in ApJ.}

\shorttitle{GU Boo: A New Low-Mass Eclipsing Binary}
\shortauthors{L\'opez-Morales \& Ribas}

\begin{document}

\title{GU Boo: A New 0.6 $M_{\sun}$ Detached Eclipsing Binary}

\author{Mercedes L\'opez-Morales\altaffilmark{1,2}}
\affil{Carnegie Institution of Washington, Department of Terrestrial
Magnetism, 5241 Broad Branch Rd. NW, Washington D.C., 20015, USA}
\email{mercedes@dtm.ciw.edu}

\and

\author{Ignasi Ribas}
\affil{Institut d'Estudis Espacials de Catalunya/CSIC, Campus UAB,
Facultat de Ci\`encies, Torre C5 - parell - 2a planta, 08193 Bellaterra,
Spain}
\email{iribas@ieec.uab.es}

\altaffiltext{1}{Carolina Royster Society Fellow 2003-2004.}
\altaffiltext{2}{Visiting Astronomer at the Southeastern Association for
Research in Astronomy (SARA) Observatory.}

\begin{abstract}
We have found a new low-mass, double-lined, detached eclipsing binary, GU
Boo, among a sample of new variables from the ROTSE-I database. The binary
has an orbital period of 0.488728 $\pm$ 0.000002 days, and estimated
apparent magnitudes $V_{\rm rotse}$ $\simeq$ 13.7 and I $\simeq$ 11.8. Our
analysis of the light and radial velocity curves of the system yields
individual masses and radii of $M_{1}$= 0.610 $\pm$ 0.007 $M_{\sun}$,
$M_{2}$= 0.599 $\pm$ 0.006 $M_{\sun}$, $R_{1}$= 0.623 $\pm$ 0.016
$R_{\sun}$, $R_{2}$= 0.620 $\pm$ 0.020 $R_{\sun}$. The stars in GU Boo are
therefore very similar to the components of the eclipsing binary YY Gem.
For this study we have adopted a mean effective temperature for the binary
of $T_{\rm eff}$ = 3870 $\pm$ 130 K. Based on its space velocities we
suggest that GU Boo is a main sequence binary, possibly with an age of
several Gyr. The metallicity of the binary is not well constrained at this
point but we speculate that it should not be very different from solar. We
have compared the physical parameters of GU Boo with current low-mass
stellar models, where we accounted for uncertainties in age and
metallicity by considering a wide range of values for those parameters.
Our comparisons reveal that all the models underestimate the radii of the
components of GU Boo by at least 10--15\%. This result is in agreement
with the recent studies of YY Gem and CU Cnc.
\end{abstract}


\section{Introduction} \label{sec:intro}

At least 70\% of the stars in the Galaxy are low-mass objects, meaning
that they have masses below 1 $M_{\sun}$. In spite of their large number,
our understanding of the physics at work in low-mass stars is still at a
basic stage. Much effort has been put in recent years into the development
of reliable models for low-mass stars (see reviews by Chabrier \& Baraffe
2000 and Allard et al. 1997). As a result of those efforts, the
state-of-the-art models now reproduce fairly well some of the parameters
of low-mass stars, such as the observed mass-$M_{V}$ and mass-luminosity
relations (e.g., Delfosse et al. 2000). However, the same models still
fail to reproduce other fundamental parameters, such as the stellar
effective temperature ($T_{\rm eff}$) scale and, most notably, the
mass-radius relation (Torres \& Ribas 2002; Ribas 2003).

As recent papers on low-mass binaries have repeatedly stated, the main
problem encountered when trying to test the models of low-mass stars is
the scarcity of accurate empirical measurements of fundamental parameters
of those objects, in particular, accurate values for both their masses and
radii. Accurate masses and radii (with errors $\sim$1--2\%) can only be
derived from double-lined, detached, eclipsing binaries (DDEBs), but until
now only three DDEBs reaching such high accuracies in their properties had
been found and analyzed in detail. Those binaries are YY Gem (Lacy 1977;
Torres \& Ribas 2002), CU Cnc (Delfosse et al. 1999; Ribas 2003), and CM
Dra (Leung \& Scheneider 1978; Metcalfe et al. 1996). Note that new masses
and radii have been recently derived for two other binaries, BW3 V38 
(Maceroni \& Montalb\'an 2004) and TrES-Her0-07621 (Creevey et al. 2005).
However, the current measurement uncertainties are too large to constrain 
the models.

We present in this work a new DDEB, GU Boo [$\alpha$(J2000) = 15:21:55.16,
$\delta$(J2000) = +33:56:04.1], composed of two similar stars of about 0.6
$M_{\sun}$ with apparent magnitudes $V_{\rm rotse}$ $\simeq$ 13.7 and
I$\simeq$ 11.8, and an orbital period close to 0.5 days. The two new
masses and radii from GU Boo increase by 40 percent\footnote{In the
analysis of Torres \& Ribas (2002), the two stars in YY Gem were assumed
to be identical. Therefore they count only as one object for statistical
purposes.} the current number of accurate measurements of those parameters
for stars below 1 $M_{\sun}$, providing two new key empirical points to
test the models.

GU Boo was first reported as a variable object by Diethelm (2001), who
listed it as a new {\it eclipsing Algol system} found among light curves
from phase I of the Robotic Optical Transient Search Experiment (ROTSE-I;
Akerlof et al. 2000). There is no reference to this binary in the
literature previous to Diethelm's identification, and it only appears in
the Guide Star Catalog (Lasker et al. 1996) as star GSC 02566-00776, and
as ROTSE1J152155.16+335604.1 in the ROTSE-I database (Akerlof et al.
2000). The variable star name $V^{*}$ GU Boo was assigned to this system
by Kazarovets et al. (2003), after Dielthelm's identification.

We identified the system as a low-mass binary based on the short orbital
period reported by Diethelm and on the estimated width of the eclipses
from the ROTSE preliminary light curve. Using the orbital period, the
width of the eclipses, and Kepler's Third Law we estimated a preliminary
total mass for the system of $M_1+M_2$ $\leq$ 1.3 $M_{\sun}$. This way of
estimating the total mass of binaries is only valid for systems with q
$\simeq$ 1.0, and close to edge-on (i $\simeq$ 90$^{\circ}$) orbital
inclinations. In any other case, the estimated $M_1+M_2$ represents only a
lower limit to the actual total mass of the binary. We subsequently
obtained complete R- and I-band light curves and radial velocity curves of
the binary to derive its masses and radii with enough accuracy to test the
models. In the following sections of this paper we report the physical
properties of GU Boo that have resulted from our detailed 
analysis of the light and
radial velocity curves, and we compare those properties to the current
models.

\section{Mean Effective Temperature} \label{sec:teff}

The temperature scale of the lower main sequence is yet weakly
established, the reason being the lack of an statistically significant
sample of low-mass stars with reliable bolometric fluxes, distances, and
radii. The increasing number of eclipsing binaries and nearby stars
resolved interferometrically should soon put a solution to this
long-standing problem. In the mean time, to determine the temperature of
M-type stars we need to rely on the use of various empirical and
model-dependent calibrations, which are usually built upon one or more
photometric indices.

We have carried out mean temperature estimations of GU Boo using several
available calibrations. Empirical and model-dependent calibrations have
been used separately to have a better measure of possible systematic
errors. In Table \ref{tab:Mlit} we summarize the magnitudes of GU Boo
collected from the literature, and used to derived the color indexes
listed in Table \ref{tab:Teff}. The first two columns of Table
\ref{tab:Teff} summarize the calibrations and color indexes that we used
to estimate the mean $T_{\rm eff}$ of GU Boo. The values of the effective
temperatures resulting from each calibration are listed in column 3.

The average temperature of GU Boo from the empirical calculations is 3900
$\pm$ 120 K, while the one obtained from the model-dependent calibrations
is 3830 $\pm$ 135 K. In all cases the temperature resulting from the (V-I)
index is generally some 200 K lower than that obtained from the (V-K)
index.  This may be indicative of some inaccuracies in the photometry. The
relatively large error bar in the resulting temperatures is a reflex of
such inconsistencies.

The average effective temperatures derived from each set of calibrations
(empirical and model-dependent) agree within the estimated errors. Given
this good agreement, we decided to take as the mean temperature of GU Boo
the average of both results. The mean effective temperature of GU Boo
adopted in our study is therefore 3870 $\pm$ 130 K. Note that the mean
temperature of GU Boo is very close to the temperature of YY Gem
determined by Torres \& Ribas (2002). This is not surprising since the
optical/IR colors of the two systems are quite similar.

As a caveat to this temperature estimation, note that we have used color
indexes collected from the literature, for which we lack information about
the epochs of each observation. Some of the magnitudes (colors) could have
been collected during eclipse epochs, and therefore the temperatures
derived from those colors would deviate from the real value of $T_{\rm
eff}$. The colors will be also affected if the magnitudes were measured at
epochs when the binary was undergoing different levels of activity.  Note
also that the V magnitude of GU Boo is an approximate calibration ($V_{\rm
rotse}$) from the ROTSE-I database (Akerlof et al. 2000)\footnote{The
ROTSE-I instrumental magnitudes $m_{\rm rotse}$ have been calibrated to
set their zero point onto a V-band equivalent scale. The recalibration
procedure is explained in detail in Akerlof et al. (2000).}. BVRI and JHK
calibrated photometry of GU Boo is currently underway at specific
out-of-eclipse orbital phases, and we expect to derive from this
photometry more accurate colors that will provide a more accurate
estimation of the mean $T_{\rm eff}$ of GU Boo.

\section{Radial Velocity Curve Observations and Analysis} \label{sec:rv}
\subsection{Observations} \label{sec:rvob}

We collected spectra covering the entire orbital phase space of GU Boo in
two nights, May 9-10 and May 18-19, 2003, with the echelle spectrograph at
the 4-m Mayall telescope at Kitt Peak. The spectra from the first night
covered a wavelength range of 310--860 nm per frame, with an average
resolving power of $\lambda$/$\Delta$$\lambda$ = 18,750 at 6000 \AA. The
configuration of the spectrograph on the second night was slightly
different (different cross-disperser) to accommodate the needs of a
parallel observing program. The resolving power of the new configuration
was the same as in night 1, but with a larger wavelength coverage,
310--1010 nm. We collected a total of 103 spectra of GU Boo in those two
nights, with average S/N between 6.5 and 9.0 (exposure times between 120
and 180 seconds). We also collected spectra of the M dwarfs GJ 410 and GJ
361 with each spectrograph configuration, to use them as templates in the
subsequent analysis of the radial velocities of GU Boo. The average S/N of
the spectra of those templates are, respectively, 82.5 and 67.0.

\subsection{Analysis} \label{sec:rvan}

Radial velocities were extracted from the spectra of GU Boo using the
implementation of the two-dimensional cross-correlation algorithm TODCOR
(Zucker \& Mazeh 1994) by G. Torres and D. Latham at the
Harvard-Smithsonian Center for Astrophysics (CfA). The analysis of GU Boo
with TODCOR was done following a procedure analogous to the one described
in detail by Torres \& Ribas (2002) in their analysis of YY Gem.

We used the spectra of the M dwarfs GJ 410 (M0~V) and GJ 361 (M1.5~V) as
trial templates instead of the using synthetic spectra, the common
procedure for more massive stars. The reasoning behind this decision is
that the synthetic spectra computed using stellar atmosphere models
reproduce well the spectral features observed in real stars down to
0.60--0.65 $M_{\sun}$ ($T_{\rm eff} \le $ 4000 K). However, for stars
below $\simeq$ 0.6 $M_{\odot}$, as is the case of GU Boo, the synthetic
spectra generated by the models deviate significantly from the ones
observed. This is because the current models do not include all the
sources of molecular opacity that become relevant below 4000 K, as is for
example the case of the H$_{2}$O molecule. To avoid the incompleteness of
the models we used instead spectra of real stars of spectral type similar
to the components of the binary.

\begin{figure}[t]
\epsscale{1.0}
\plotone{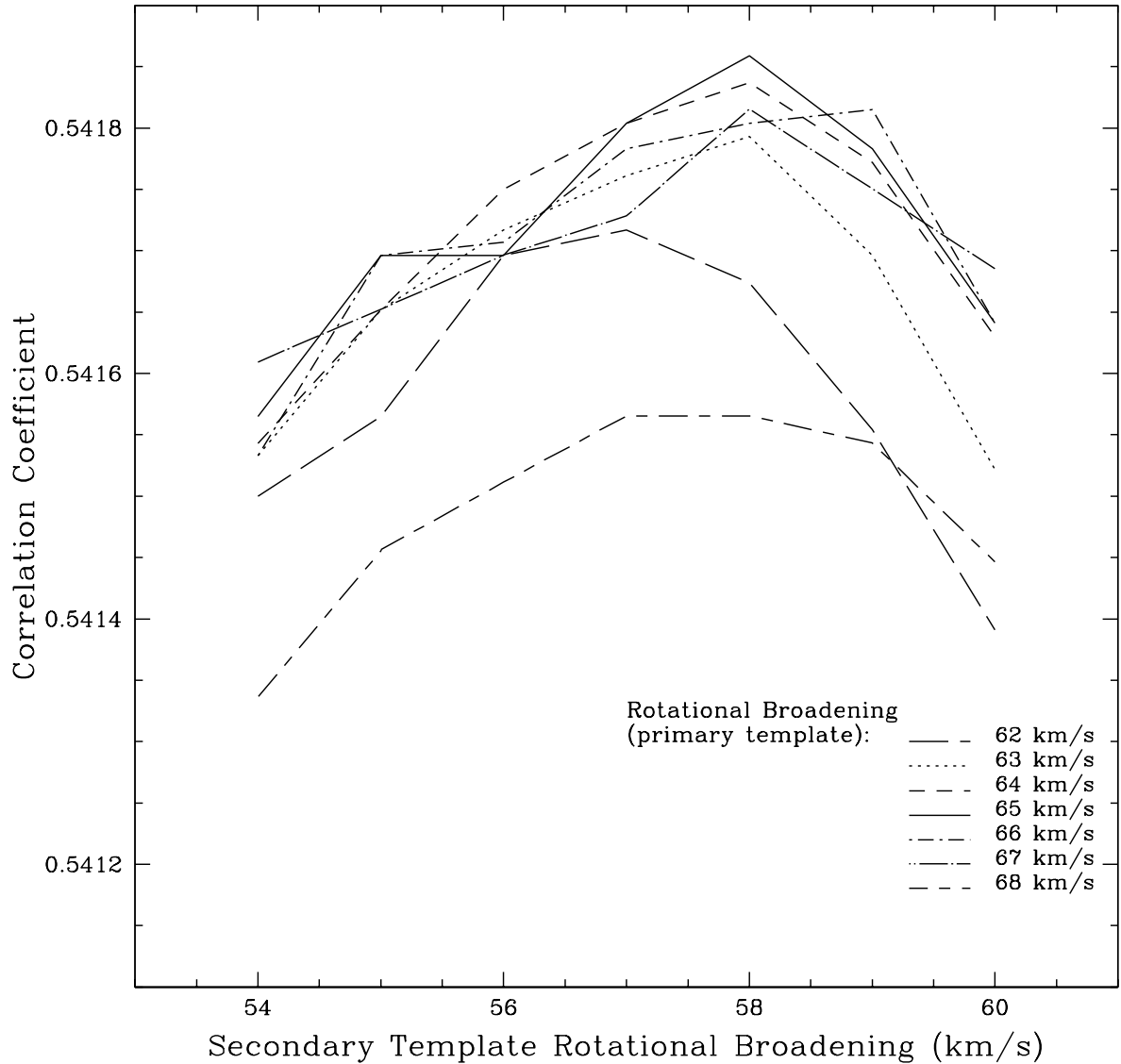}
\caption{Variation of the correlation coefficient in TODCOR as a function of
the rotational broadening of the template of the secondary. We use the spectra
of GJ 410 as template for both the primary and the secondary. Each curve in
the figure corresponds to a fixed value of the rotational broadening of the
primary template (see values in the lower-right corner). The values of the
broadening of the secondary are represented in the horizontal axis. The highest
value of the correlation coefficient in TODCOR corresponds to rotational
broadenings of 65 km~s$^{-1}$ for the primary, and 58 km~s$^{-1}$ for the
secondary.}
\label{fig:todcor}
\end{figure}

Optimal results with TODCOR are attained when the rotational broadening of
the templates matches the broadening of the components of the binary. We
artificially broadened the spectra of the templates by different amounts
of $v \sin i$, and ran extensive tests to determine the combination of
templates and rotational broadenings that best matched the spectra of GU
Boo. The results of those tests using GJ 410 as template for both stars
are shown in Figure \ref{fig:todcor}. Similar tests using GJ 361 for each
star and GJ 410 and GJ 361 for either the primary or the secondary gave
always lower correlation coefficients than the tests using GJ 410 for the
two stars. The highest value of the correlation coefficient in TODCOR
indicates the best match between the observed spectra and the combination
of broadened templates. In this case, as illustrated in the figure, the
highest value of the cross-correlation coefficient occurred when we used
GJ 410 as template for both stars, broadened by 65 km~s$^{-1}$ for the
primary and 58 km~s$^{-1}$ for the secondary.

The S/N in each order of the echelle spectra degrades towards the edges
because of the blaze function of the spectrograph. When this effect is
combined with the already low S/N of the data, the information in those
edges turns out to be too poor even for a technique as sensitive as
TODCOR. We removed the very-low S/N parts of the spectra by trimming in
each aperture the edge portions with S/N $\leq$ 3.5. We also discarded all
the apertures that had S/N below that threshold, or that contained
prominent non-stellar absorption bands or emission lines.  The radial
velocity curves were derived running TODCOR separately for each aperture
and then computing an average radial velocity per frame for each star. The
resulting radial velocities correspond to the average of all the apertures
in each frame, after eliminating the values that deviate more than
2$\sigma$ from the mean. The typical standard deviation of the average
velocities is 10 km s$^{-1}$ or better.

We also checked for possible systematic effects in the individual
velocities derived from TODCOR by performing numerical simulations. We
generated 103 artificial binary spectra by combining templates of GJ410
with different velocity shifts. The velocity shifts were computed from the
best available orbital solution of GU Boo and the times when the real
observations of the binary where taken. We then ran TODCOR substituting
the real observations by the synthetic spectra to measure any systematics.
No clear systematic trend was noticed, and the observed differences
between the input and output velocity shifts were of the same order as the
average velocity dispersion per frame. Finally, we found that the changes
in the orbital solution of the radial velocity curves after correcting for
the errors suggested by the simulations were insignificant, therefore we
did not correct for these residuals in the computation of the final
spectroscopic orbital solution of GU Boo.

\begin{figure}[t]
\epsscale{1.0}
\plotone{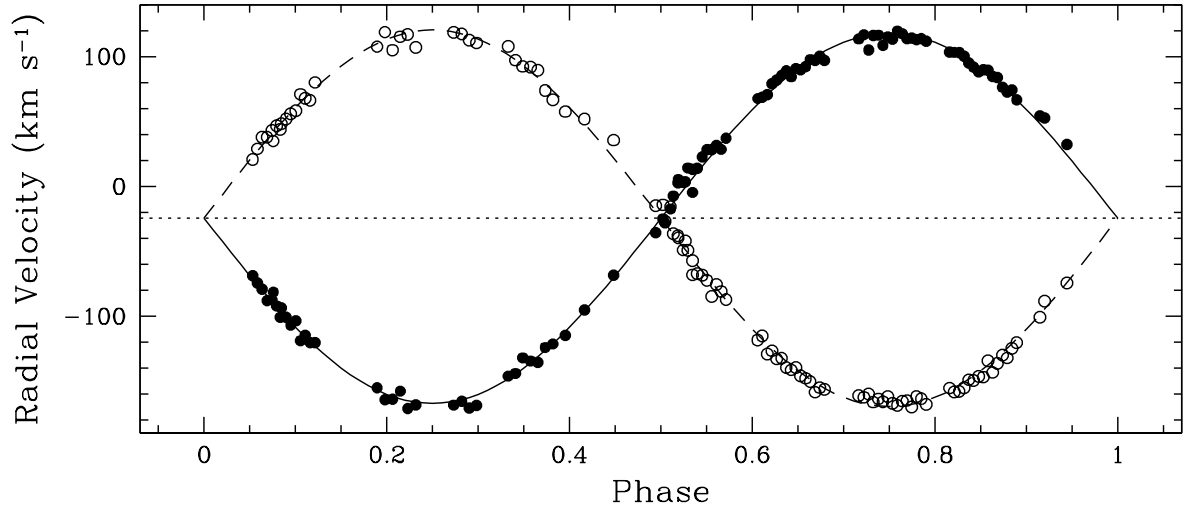}
\caption{Radial velocity curve of GU Boo. The filled and open circles
correspond to the velocities of the primary and the secondary, respectively.
The solid lines represent the orbital solution obtained with TODCOR, and the
dashed line is the velocity of the center of mass of the system.}
\label{fig:rv}
\end{figure}

The final radial velocity curves are shown in Figure \ref{fig:rv}. The
filled circles in that curve correspond to the radial velocity of the
primary, the open circles to the radial velocity of the secondary, and the
solid and dashed lines show the best orbital solution fits. Table
\ref{tab:Params} summarizes the parameters of the spectroscopic orbital
solution derived from those curves. In that solution we adopted the
orbital period P(days) and initial epoch $T_{0}$(HJD) derived in \S
\ref{sec:lcpe}. We left the eccentricity of the orbit as a free parameter
during the preliminary stages of the computation of the orbital solution
and then ran several tests using different initial values of that
eccentricity. Those tests resulted on an average orbital eccentricity of
$e$=0.0001 $\pm$ 0.0038, which suggests a circular orbit for the binary.
Based on this result, the eccentricity of the orbit of GU Boo was the set
to zero in the analysis.

Systematic errors in the radial velocities could be caused by the spots on
the surface of the stars. We evaluated this effect using the spot scenario
adopted in \S \ref{sec:lcan} and found systematic deviations in the
velocity semi-amplitudes of the order of 0.5 km s$^{-1}$. Note from the
values of $K_{1}$ and $K_{2}$ in Table \ref{tab:Params} that 0.5 km
s$^{-1}$ nearly corresponds to the 1$\sigma$ errors reported for those
parameters. We could not, however, apply the corrections to the radial
velocities because the photometric and spectroscopic observations were not
contemporaneous and possibly had different spot configurations. Thus, we
caution that systematic errors in $K_{1}$ and $K_{2}$ of the same order of
the random errors could be present.

The radial velocity data in Figure \ref{fig:rv} is available in electronic
form from the ApJ website. The contents of the electronic table are
illustrated in Table \ref{tab:RVs}. The velocities in that table have been
corrected to the heliocentric reference system by adopting a peculiar
velocity of $-$13.9 km~s$^{-1}$ for the template of GJ 410 (Gliese 1991).

\begin{figure}[t]
\epsscale{1.0}
\plotone{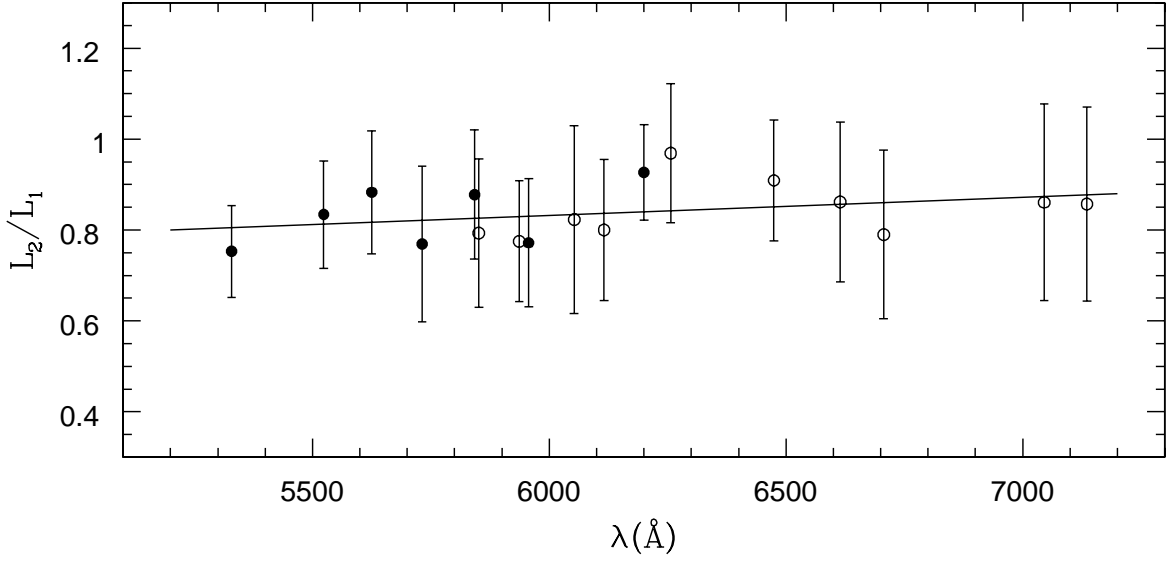}
\caption{Light ratio of the components of GU Boo estimated by TODCOR as a
function of wavelength. The filled and open circles correspond, respectively,
to the spectra collected on the first and second night of observations. The
continuous line represents the best linear fit to the data. Based on that fit,
we have adopted an average luminosity ratio for the binary of $L_{2}/L_{1}$
$\simeq$ 0.84 $\pm$ 0.04 (at 6200\AA).}
\label{fig:lrat}
\end{figure}

In addition to extracting the radial velocity semi-amplitudes of the stars
in the binary, the cross-correlation algorithm TODCOR (Zucker \& Mazeh
1994) also computes their luminosity ratio, independently from the light
curve analysis. The luminosity ratio $\alpha$= $L_{2}/L_{1}$ is
implemented in TODCOR as an additional parameter to account for the
relative intensity of the spectrum of each star. The value of $\alpha$ can
be fixed by the user or left free, letting the algorithm compute it as the
value that maximizes the correlation parameter in TODCOR for a set of
templates. We estimated the luminosity ratio of GU Boo as a function of
wavelength by leaving $\alpha$ as a free parameter, and fixing the rest of
the parameters in TODCOR to the values used to derive the spectroscopic
orbital solution. The results are presented in Figure \ref{fig:lrat}. The
relatively large dispersion of the results is a consequence of the low S/N
of the data. We applied a linear fit, represented by the continuous line
in the figure, which yields an average value of the luminosity ratio of
$L_{2}/L_{1}$ $\simeq$ 0.84 $\pm$ 0.04 at 6200 \AA. We believe that the
slight slope of the fit is real, and that it is caused by the difference
in temperature of the two components.

\section{Light Curve Observations and Analysis} \label{sec:lc}
\subsection{Observations} \label{sec:lcob}

We obtained complete R- and I-band light curves of GU Boo using standard
CCD differential photometry techniques. The data were collected over 7
nights between March 24 and May 6 2003, at the Southeastern Association
for Research in Astronomy (SARA) 0.9-m telescope at Kitt Peak. The average
photometric accuracy of individual measurements is 0.003 mag in R and
0.004 mag in I, respectively.

The SARA telescope was equipped with an Ap7 512$\times$512 CCD that
provided a field-of-view of 6$\arcmin$$\times$6$\arcmin$. GU Boo is a
relatively faint target ($V_{\rm rotse}$ $\simeq$ 13.7), with not many
other objects of similar spectral type or brightness nearby. By placing GU
Boo close to one of the corners of the CCD, we managed to strategically
locate the binary on the chip together with two other stars of similar
apparent magnitude. As comparison star we selected GSC 02566-00631,
located 2$\farcm$59 away from the target. The check star was chosen to be
GSC 02566-00935, at about 4$\farcm$53 from GU Boo, and 1$\farcm$94 from
the comparison star. Both stars passed respective tests for intrinsic
photometric variability and proved to be stable during the time span of
our observations.

We collected a total of 492 measurements in the R band and 947 in the I
band with exposure times of 120 and 60 seconds, respectively. The
observations covered the entire orbit of the binary in each filter. The
data were analyzed using standard aperture photometry packages in IRAF,
with no differential extinction effects taken into account given the
relative small separation between the target and the comparison and check
stars.

The R and I light curves of GU Boo that resulted from including all the
data above presented significant scatter in brightness (of up to 0.1 mag)
over all orbital phases. The apparent time scale of those brightness
variations is of the order of days to weeks, and we attribute them to
migration of spots on the surface of the stars. However, the photometry is
fairly stable within a given night, with most of the scatter appearing at
orbital phases when observations made several days or weeks apart overlap.
The photometric dispersion of the light curves was substantially reduced
by only using the data from nights that (i) when combined cover most parts
of the orbital phases of the system, and (ii) were collected close enough
in time for the effect of spot migration to be minimal, so the spots can
be satisfactorily modeled.

\subsection{Orbital Period and Ephemeris} \label{sec:lcpe}

The first to provide an orbital period and zero epoch for GU Boo was
Diethelm (2001). Diethelm, using the preliminary light curve from ROTSE-I
(Akerlof et al. 2000), measured an orbital period of P= 0.488722 $\pm$
0.000002 days, and established the initial epoch $T_{0}$(HJD)=2451222.8497
$\pm$ 0.0012 for the occurrence of the primary eclipses of the system.
However, given that the quality of the ROTSE-I light curve used by
Diethelm is rather poor, we decided to check his results using the
ephemeris from our new light curves. Our data covers seven different
eclipse epochs, including four primary and three secondary minima.
Accurate times for those minima were computed by applying 6th order
polynomial fits to the light curve data during the eclipses. The values
that we obtained are listed on the third column of Table \ref{tab:Tmin}. A
linear least squares fit to the four primary minima in that table yields
an orbital period of P= 0.488728 $\pm$ 0.000002 days, which is in
acceptable agreement with the value found by Diethelm. As new reference
epoch we have adopted the first time of minimum of the primary eclipse in
our data, i.e. $T_{0}$(HJD)=2452723.9811 $\pm$ 0.0003. The re-derived
ephemeris equation of the binary is therefore:
\begin{equation}
T(\mbox{Min I}) = \mbox{HJD}2452723.9811(3) + 0.488728(2) \cdot E
\end{equation}

The (O-C) residuals for each time of minimum, i.e. the difference between
the observed times of minima and the ones predicted by the ephemeris
equation above, are listed in the last column of Table \ref{tab:Tmin}. The
average residuals are 0.00000 $\pm$ 0.00037 days and 0.00030 $\pm$ 0.00029
days for the primary and secondary minima, respectively. Using the new
orbital period derived above we find an average phase difference between
primary and secondary eclipse minima of $\Delta\phi$ = 0.49981 $\pm$
0.00033, which suggests a circular orbit.

\subsection{Analysis} \label{sec:lcan}

\begin{figure}[t]
\epsscale{1.0}
\plotone{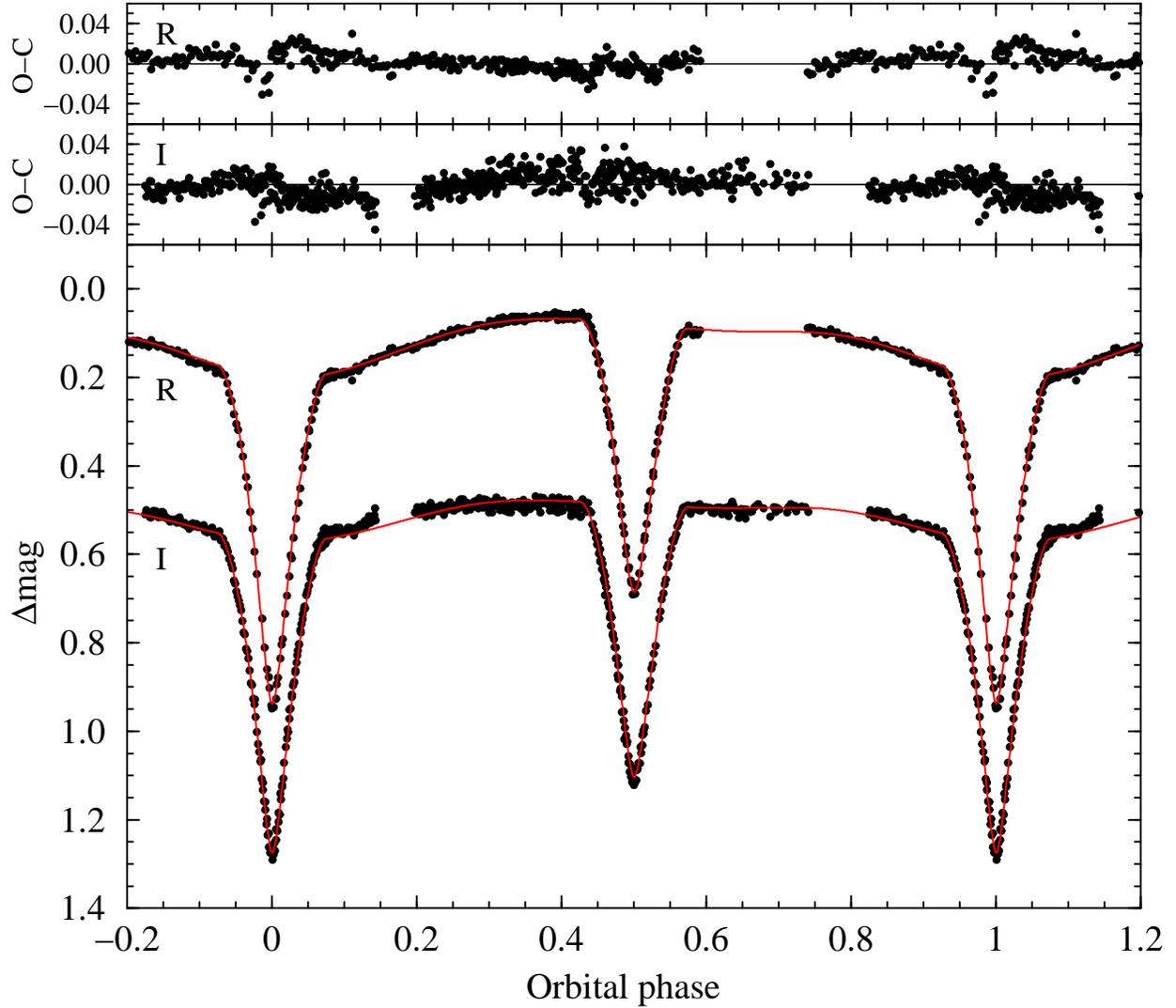}
\caption{R- and I-band light curves of GU Boo collected at the SARA 0.9-m
telescope in KPNO, between March 24 and May 6 2003. The dots represent
individual observations in each passband, and the continuous lines correspond
to the best fits to those light curves using WD2003 (see \S \ref{sec:lcan} for a detailed
discussion of the fitting process). The two diagrams on the top show the
residuals of those fits. The average values of the residuals are 0.0086 mag in
R and 0.0118 mag in I, respectively.}
\label{fig:lcs}
\end{figure}

The final light curves used in our analysis contain 365 observations in R
and 622 in I and are shown in Figure \ref{fig:lcs}. Table
\ref{tab:PhotNights} lists the dates when the data in Figure \ref{fig:lcs}
were collected, the range of orbital phases covered by those observations,
and the number of data points per phase interval.  A few sample R-band
light curve measurements are listed in Table \ref{tab:LCsR}. The full R
and I--band photometry tables are available in electronic form from the
ApJ website.

Both R and I light curves were fitted simultaneously using the May 2003
version of the Wilson-Devinney program (Wilson \& Devinney 1971, May 2003
revision; WD2003 hereafter). We fixed beforehand as many of the free
parameters in WD2003 as possible to reduce computing time and convergence
problems. The values assigned to those parameters are based on information
from external sources or from preliminary fits to the light curves (see
column 5 in Table \ref{tab:PhotSols}).

We assigned to GU Boo a detached binary configuration after ruling out the
contact binary scenario, since the estimated ratio of the stellar radii to
the Roche lobe radius is $r_{*}/r_{\rm rl} \simeq 0.58$. For completeness,
we also included reflection and proximity effects in the fits, although
there is no clear evidence for such effects in the light curves.

Wilson-Devinney takes into account the effect of limb darkening, gravity
brightening, and bolometric albedo when computing the fluxes of the stars.  
To model the limb darkening we chose a square-root law like the one
proposed by D\'{\i}az-Cordov\'es \& Gim\'enez (1992). The first and second
order limb darkening coefficients were interpolated from tables computed
from the Phoenix atmosphere models (e.g., Allard \& Hauschildt 1995).
Since the temperatures may change during the fitting process, the limb
darkening coefficients were recomputed for each iteration of WD2003
whenever necessary. The gravity brightening coefficients were set to 0.2
for both stars, based on the values proposed by Claret (2000). Finally we
set the bolometric albedo coefficients to 0.5, which is the standard value
for stars with fully convective envelopes, as is the case of the
components of GU Boo.

We used the orbital solution of the radial velocity curves (\S
\ref{sec:rvan}) to set the mass ratio of the binary to $q=M_2/M_1=0.9832$,
and the eccentricity of its orbit to zero (although we ran some initial
fits to the light curves to further confirm the circularity of the orbit).
We also conducted a number of tests to evaluate the presence of any third
light ($L_3$) in the light curves. Those tests yielded null results. The
orbital period and the initial epoch of GU Boo were adopted as
$P=0.488728$ days and $T_0(\mbox{HJD})=2452723.9811$, respectively, using
the values of the ephemeris derived above. We also assumed that the stars
are rotating synchronously. This assumption later proved to be consistent
with the radii and orbital inclination resulting from the fit and with the
rotational velocities of the stars estimated from the rotational
broadening of the spectra (\S \ref{sec:rvan}).

After setting all the parameters above we ran WD2003 iteratively to find
the best solution for the remaining light curve parameters. The parameters
left free were a phase shift $\phi$, the orbital inclination of the binary
$i$, the temperature of the secondary $T_2$, the surface potential of each
star, $\Omega_1$ and $\Omega_2$, the luminosity of the primary $L_1$, and
the spot parameters. We assigned the same statistical weight to each
individual observation within a light curve. As relative weight between
light curves we used the mean standard deviation of the measurements in
each curve (0.003 mag in R and 0.004 mag in I, respectively). The WD2003
iterations were run automatically and a solution was defined as the set of
parameter values for which the differential corrections suggested by the
code were smaller than the statistical errors for three consecutive
iterations (e.g., Kallrath \& Milone 1999). We ran a number of fits with
different starting values to explore the full parameter space and to avoid
missing the global minimum for a local minimum.

The initial tests indicated that the convergence on most of the free
parameters was very fast. However, we soon realized that the optimization
of spot properties would highly complicate the analysis. Widely differing
spot configurations can lead to light curve fits of almost identical
quality \footnote{The ambiguity of photometric solutions for spotted
binaries is a well-known problem and was one of the reasons to develop
Doppler imaging techniques (see, e.g., Bell et al. 1990; Maceroni \& Van't
Veer 1993;  van Hamme et al. 2001).}.  The intrinsic properties of the
binary components are seldom noticeably affected by the adopted spot
parameters and, quite often, reaching a good fit to the out-of-eclipse
variations can be regarded as an aesthetic exercise. GU Boo turned out to
be somewhat of an exception to this general behavior. We found that
different spot parameters do have a rather sizeable effect on the physical
properties of the components. Thus, careful modeling of the stellar spots
on the components of GU Boo is indeed a critical issue.

We explored a wide range of spot scenarios in an attempt to reproduce the
out-of-eclipse brightness variations of the light curves. We also carried
out tests to evaluate the sensitivity of the light curve to variations of
each spot parameter (latitude, longitude, angular radius, temperature
relative to the photosphere, i.e. temperature factor). The fits turned out
to be marginally better when the spots have central latitudes in the range
$30-45^{\circ}$. Our result can be compared with the theoretical
calculations of Granzer et al. (2000) who found that the most likely
latitude location for spots on the surface of 0.6~M$_{\sun}$ main sequence
stars is $\sim45^{\circ}$. Our tests also revealed that the fitting
procedure was extremely unstable when more than two spot parameters were
left to vary. To decouple this correlation we ran WD2003 using the
so-called Method of Multiple Subsets (MMS). MMS allows parameters to be
fitted on separate subsets, forcing strongly correlated parameters to
converge independently.

We ran series of fits for different number of spots, temperature factors,
longitudes, and angular sizes. Our tests started with one dark spot (i.e.
temperature factor smaller than 1) on either component. The fit in this
case was better than the solution with no spot but the residuals still
showed strong systematic deviations.  Thus, we tried a series of solutions
with two dark spots. Out of the four possible combinations, a good fit was
only achieved when the two spots were located on the primary
component\footnote{The primary has been defined as the star eclipsed at
orbital phase $\phi=0.0$ in the light curves.} The solutions with one dark
spot on each component and two dark spots on the secondary yielded
significantly poorer fits and were discarded. The configuration with two
spots on the primary reproduces reasonably well the features observed in
the light curves and results in root mean square residuals of 0.0087 mag
and 0.0115 mag in R and I, respectively. Column 3 in Table
\ref{tab:PhotSols} gives a summary of the light curve parameters resulting
from this fit (spot scenario \# 1). This solution, however, is
unsatisfactory for a number of reasons: (i) the predicted luminosity
ratios in each passband are significantly larger
($L_2/L_1\approx0.90-0.96$) than the values computed from the spectra of
GU Boo (see \S \ref{sec:rvan} and Figure \ref{fig:lrat};
$L_2/L_1\approx0.82-0.85$), (ii) the radius of the secondary (less
massive) star is larger than the radius of the primary, which would be
inconsistent with GU Boo being a main sequence system (see \S
\ref{sec:age}), and (iii) given the similarity in mass of the two
components, it is difficult to explain how one star could be heavily
spotted while the other one completely immaculate.

\begin{figure}[t]
\epsscale{1.0}
\plotone{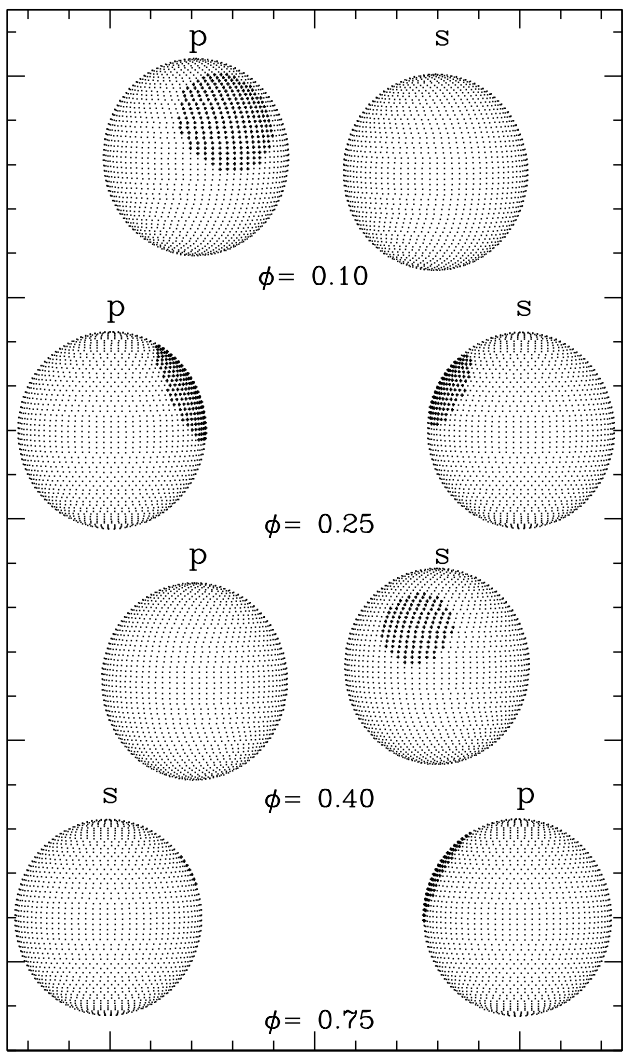}
\caption{Representation of the spot configuration of GU Boo at various orbital
phases. The primary and secondary stars are labelled as {\it p} and {\it s},
respectively. The spot on the primary is cooler than the photosphere, with a
temperature factor of $T_{sp}/T_{ph}$ = 0.94. The spot on the secondary is
a ``bright spot'' (hotter than the photosphere), with  $T_{sp}/T_{ph}$ =
1.06. See Table \ref{tab:PhotSols} for a summary of the parameters of the
spots.} 
\label{fig:spots}
\end{figure}

\begin{figure}[t]
\epsscale{1.0}
\plotone{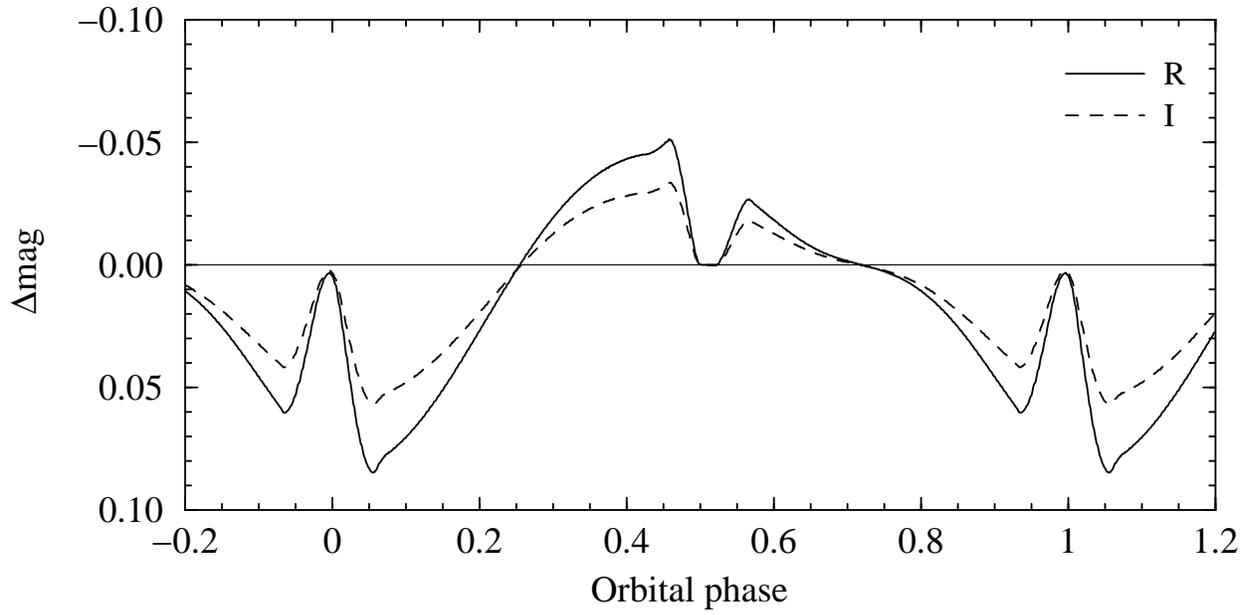}
\caption{Photometric effect of the spots on the light curves. The continuum and
dashed lines correspond, respectively, to the R- and I-band curves.}
\label{fig:eff}
\end{figure}

We tried some alternative scenarios to find a more physically realistic
solution.  Adding more dark spots did not improve the quality of the fits.
We then considered a configuration with two spots but one of them being
brighter than the surrounding photosphere (i.e. temperature factor greater
than 1.0). After several tests we arrived at a configuration that yielded
a fit of very similar quality to the two dark spot scenario described
above. The root mean square residuals in this case are respectively 0.0086
mag in R and 0.0118 mag in I. The resulting orbital solution parameters
are listed in column 4 of Table \ref{tab:PhotSols} (spot scenario \# 2).
This second fit is much more satisfactory because: (i) the luminosity
ratios in the two passbands are in good agreement with the independent
values found from the spectroscopic analysis, (ii) the radii of the stars
are very similar and compatible with the expected ratio from a main
sequence evolutionary stage, and (iii) the model predicts surface
inhomogeneities (spots) in both stars. The resulting light curve fit is
shown in Figure \ref{fig:lcs}. A graphical representation of this spot
configuration is shown in Figure \ref{fig:spots}. The photometric effects
of the spots on the light curves are illustrated in Figure \ref{fig:eff}.

Being more physically valid, the only feature of this solution that needs
further explanation is whether the presence of bright spots on the
photosphere is realistic. This is not the first time that bright spots
(plages) are suggested in low-mass binaries. Studies of YY Gem by Torres
\& Ribas (2002), Butler \& Doyle (1996), and Kron (1952) mention possible
{\it bright regions} to explain features observed in the light curves of
that binary. Another case is that of the contact binary YY Eri (Maceroni
et al. 1994). Moreover, a low-mass binary currently being analyzed by us
(RXJ 0239.1-1028) shows indications of bright areas on the surface of the
stars as well. We would like to mention, however, that bright spots on
stellar surfaces can have an alternative explanation when only photometric
data are being modeled. A light curve fitting program like WD2003 has a
very simple spot model (circular, homogeneous spots) that is only able to
measure the contrast between different areas on the surface of the star.
The model of a star with a bright region at a certain location is
equivalent to a model of the star with a dark area, of proportional
temperature factor, covering the complementary surface. Thus, our scenario
can be equivalently interpreted as a more or less uniform distribution of
dark spots covering most of the stellar surface except for a spot-free
area. This spot-free area can be modeled as a ``bright spot'' while, in
reality, it represents the true photosphere. Fortunately, all this has no
effect on the physical stellar properties but is just a reflex of the
inherent indeterminacy of the spot models. It is only with future Doppler
tomography studies of these binaries that the open question about the
presence of bright surface homogeneities can be more conclusively
addressed. In the meantime, we have decided to adopt this second scenario
with a bright spot because it yields physically realistic parameters.

Note also that our model calls for the presence of large spots covering a
significant fraction of the stellar surfaces. The presence of such large
spots is not a surprising result, given the high level of activity
characteristic of low-mass stars. But a more realistic picture would be
that of smaller groups of spots with larger temperature contrast
distributed over the same surface area of the star covered by the single
spot in the model (this has sometimes been referred to as a ``starpatch'';
see discussion in Toner \& Gray 1988). In average, those groups of smaller
spots would produce the same modulation in brightness as a single large
spot with less contrast in temperature. The only difference is that an
inhomogeneous distribution of small spot groups will cause rapid
fluctuations in brightness as the spotted star undergoes eclipse and parts
of the stellar surface with and without spots are gradually occulted. This
kind of fluctuations is probably what we observe as larger residuals to
the fits around the eclipses in Figure \ref{fig:lcs}. Future very accurate
photometry (for example from upcoming space missions) will improve the
current coarse spot modeling by permitting detailed eclipse tomography.

Finally, there is a detail in Table \ref{tab:PhotSols} that deserves
further discussion, which is the calculation of the stellar temperatures.
The temperatures used by WD2003 correspond to the defined stellar
(immaculate) photosphere. In cases where the spot areal coverage is small,
this is a good representation of the overall effective temperature (i.e.,
related to the bolometric luminosity). However, in our case, spots cover a
significant fraction of the stellar surfaces and thus the working
temperatures in WD2003 deviate significanty from the stellar effective
temperatures. We have therefore devised a procedure to calculate the
effective temperature of a star with a large, either dark or bright, spot
from its immaculate photospheric temperature. It is easy to demonstrate
that the surface covered by a spot that subtends an angle of $2\theta_{\rm
sp}$ on a sphere of radius $R$ is:
\begin{equation} 
S_{\rm sp}=2 \pi R^2 (1-\cos \theta_{\rm sp})
\end{equation} 
Then, the effective temperature ($T_{\rm eff}$) is calculated from 
the photospheric temperature ($T_{\rm ph}$) and the spot temperature
($T_{\rm sp}$) using the areal coverage as the weighing factor:
\begin{equation} 
{T_{\rm eff}}^4 = \frac{S_{\rm sp} {T_{\rm sp}}^4 + S_{\rm ph} {T_{\rm 
ph}}^4}{S_{\rm tot}} 
\end{equation} 
This can be parameterized as a function of $\kappa$, which is the
temperature factor ($T_{\rm sp} = \kappa T_{\rm ph}$), to obtain:
\begin{equation} 
T_{\rm eff} = \left( 1-\frac{(1-\kappa^4)}{2} (1-\cos \theta_{\rm sp}) 
\right)^{1/4} T_{\rm ph} 
\end{equation} 
Using this expression we have corrected the working photosphere
temperatures from WD2003 (in Table \ref{tab:PhotSols}) to compute the
effective temperatures for the components. To arrive at the final values
\ref{tab:AbsDim} we adjusted the individual temperatures to yield the mean
system temperature computed in \S \ref{sec:teff}.

\section{The GU Boo System} \label{sec:system}
\subsection{Absolute Parameters of the Components} \label{sec:params}

We used the parameters of the photometric and spectroscopic orbital
solutions derived in the previous two sections to compute its absolute
dimensions and physical parameters. The results are summarized in Table
\ref{tab:AbsDim}. The two stars in GU Boo are very similar, as their mass
and radius ratios indicate. The ratio of the masses is $M_{2}/M_{1}$=
0.9832 $\pm$ 0.0069, while the ratio of the radii is $R_{2}/R_{1}$= 0.99
$\pm$ 0.04. The shape of the stars is practically spherical, with their
radii being only $\simeq$ 3\% larger at the equator than at the poles (see
values of the fractional radii in Table \ref{tab:PhotSols}). This small
elongation is caused by a combination of tidal interaction between the
stars and rotational flattening effects.

The radius of each star is consistent with GU Boo being a main sequence
system (see \S \ref{sec:age}), and the larger error bars in $R_{2}/R_{1}$
result from the complications that the effects of spots introduce in the
modeling of the light curves. Upcoming JHK photometry of GU Boo will give
a second independent estimation of the radii of the stars by providing
light curves GU Boo under a different spot configuration and at
near-infrared wavelengths, where the effect of spots diminished with
respect to the visible.

We derived the values of the projected rotational velocities of the stars
from the orbital period of the binary (assuming synchronous rotation), and
the radii of the stars in Table \ref{tab:AbsDim}. The resulting projected
rotational velocity of the primary, ${v_{\rm sync}}_1 \sin i$= 64.4 $\pm$
1.6 km s$^{-1}$, agrees with the value estimated using TODCOR. In the case
of the secondary, the resulting velocity ${v_{\rm sync}}_2 \sin i$= 64.1
$\pm$ 2.1 km s$^{-1}$ is 10\% larger than the value of the best estimation
from TODCOR. However, this discrepancy is consistent with the
uncertainties in the radii that result from the fits to the light curves.
We recomputed the radial velocities in TODCOR replacing the secondary
template by a new template broadened by 64 km s$^{-1}$ to find a variation
of only 0.33\% in the cross-correlation coefficient and no appreciable
differences in the parameters of the binary derived from either radial
velocity curve.

The relative difference in temperature of the two stars (${T_{\rm
eff}}_2/{T_{\rm eff}}_1$= 0.972) is consistent with the differences
between their masses and between their radii. However, a more accurate
determination of the mean absolute temperature of GU Boo is still
necessary. The value adopted for the temperature of the primary in this
work, ${T_{\rm eff}}_1$ = 3920 K, is a preliminary estimation based on the
average of the temperatures computed in \S \ref{sec:teff}. Note that the
potentially large uncertainty in the effective temperature of the binary
has no impact on the accuracy of the determined absolute dimensions. For
example, the radii of the stars, which are obtained from the light curve
modeling, suffer inappreciable changes when $T_{\rm eff}$ values $\pm$300
K (2$\sigma$) about the mean are adopted.

The luminosity and absolute bolometric magnitudes $M_{bol}$ of the stars
in Table \ref{tab:AbsDim} were computed from their effective temperatures
and their radii. We applied bolometric corrections $BC_{V}$ to the
$M_{bol}$ of each star to derive their absolute visual magnitudes $M_{V}$.
Those bolometric corrections were computed using the models in Table
\ref{tab:Teff}. The values obtained for each star are respectively,
$BC_{V_{1}}$= $-$1.14 $\pm$ 0.08 mag and $BC_{V_{2}}$= $-$1.29 $\pm$ 0.08
mag. Finally, we have estimated a distance of d $\simeq$ 140 $\pm$ 8 pc
for GU Boo from the distance modulus equation, using the joint absolute
magnitude of the system $M_{V}$= 7.98 mag and assuming $m_{V}$=$m_{\rm
rotse}$= 13.7 mag.

\subsection{Age and Space Velocities} \label{sec:age}

An estimate of the age of GU Boo is an important parameter for a critical
evaluation of stellar models. Such age estimation, however, is far from
straightforward for low-mass stars because of their characteristic long
evolution timescale. It is only during the pre-main sequence stage, which
lasts for a few $10^8$ yr (Chabrier \& Baraffe 1995; Palla \& Stahler
1999), that the properties of an M-type star suffer appreciable changes.
Once on the main sequence, a stars with the masses of GU Boo will evolve
at a very slow pace for tens of Gyr. Therefore, in our case, the key task
is to find out whether the components of GU Boo could be in the pre-main
sequence stage or if they have already arrived on the main sequence.

GU Boo appears to be an isolated system, not associated to any known
cluster, stellar association, or star formation region. Therefore, a
kinematic age estimate is the only possibility to evaluate whether the
binary belongs in a young or old population. The relevant quantities for
this study are the three components of the space velocity
$(U,V,W)$\footnote{According to our convention, positive values of $U$,
$V$, and $W$ indicate velocities towards the galactic center, galactic
rotation and North galactic pole, respectively.}. The heliocentric space
velocity components of GU Boo were computed from its position, radial
velocity ($\gamma=-24.57\pm0.36$ km~s$^{-1}$; see Table \ref{tab:Params}),
distance ($d=140\pm8$ pc) and proper motions. The latter were retrieved
from the USNO-B1.0 catalog (Monet et al. 2003): $\mu_{\alpha}=26\pm4$ mas
yr$^{-1}$; $\mu_{\delta}=-28\pm1$ mas yr$^{-1}$. The resulting space
velocity components are: $U=14.3\pm1.8$ km~s$^{-1}$; $V=-12.6\pm1.5$
km~s$^{-1}$; $W=-28.1\pm1.3$ km~s$^{-1}$, which correspond to a total
space velocity of $S=34.0\pm2.7$ km~s$^{-1}$.

We can infer from those space velocities whether GU Boo is a young
population system (age $<300$ Myr), which would imply a pre-main sequence
stage, or if it has, on the contrary, parameters characteristic of an
older population. There are a number of indications that favor the latter
case and thus suggest that GU Boo has reached the main sequence: {\it 1)}
The location of the star on the $UV$ plane does not fall within the area
(although it lies not far from it) defined by Eggen (1984, 1987) as
corresponding to the young disk population, nor in the area covered by
young population tracers (Skuljan et al. 1999); {\it 2)} The space
velocities do not match those of any known moving group (Montes et al.
2001); {\it 3)} The modulus value of the $W$ velocity component
(perpendicular to the Galactic plane) is significantly larger than
expected for a young star ($W\approx-7\pm10$ km~s$^{-1}$; e.g. Mihalas \&
Binney 1981; Nordstr\"om et al. 2004), suggesting that GU Boo has been
subject to ``disk heating'' processes, which operate over timescales
longer than the galactic rotation period (200--250 Myr). There is always
the possibility, however, of GU Boo being a young runaway object that has
been ejected at relatively high velocity from its birthplace. In this
case, its velocity would not correspond with its kinematic age. We cannot
exclude this scenario with the current data but there is neither any
indication favoring it. From the available evidence, we suggest that GU
Boo is not a very young object and may possibly have an age of several
gigayears. In addition, we do not expect the metal abundance of GU Boo to
differ very significantly from the solar one, since the space motions are
compatible with a disc population.

\subsection{Stellar Activity Indicators} \label{sec:activity}

We provide here a comparison between some stellar activity indicators in
GU Boo, and the same activity indication parameters observed in YY Gem and
CU Cnc. In particular we focus on the $H\alpha$ and X-ray emission
observed in the three objects. In the spectra of GU Boo collected to
derive the radial velocity curves we observed clear evidences of strong
emission in $H\alpha$. The observed emission features were double-peaked,
and reached a maximum value of the continuum corrected equivalent widths
of $\simeq$ 1.7$\pm$0.1 \AA\ at orbital phase $\phi$ $\simeq$ 0.01 (around
primary eclipse). There was no apparent correlation between the phase of
the maximum $H\alpha$ emission in the spectra and the phase of the spots
in the light curves, although the reason for this is that each dataset was
collected in a different epoch. The observed levels of $H\alpha$ emission
in GU Boo are lower than the emission levels in YY Gem ($\simeq$ 2 \AA)
and CU Cnc (3.85 \AA\ and 4.05 \AA) reported by Young et al. (1989) and
Ribas (2003), respectively.

GU Boo also appears in the ROSAT All-Sky Bright Source Catalog (Voges et
al. 1999), as X-ray source 1RXS J152155.7+335625. Using the calibration
equation by Schmitt et al. (1995) we have estimated an X-ray luminosity of
$\log L_{X}$ (ergs s$^{-1}$)= 29.3$\pm$0.2 for GU Boo, which is very
similar to the X-ray luminosity of YY Gem, $\log L_{X}$ (ergs s$^{-1}$)=
29.27$\pm$0.02, computed using the same calibration equation and the
parameters of this binary from ROSAT. This result indicates that the
levels of X-ray activity for both systems are very similar. Furthermore,
the fact that the estimated X-ray luminosity of GU Boo agrees with that of
YY Gem indicates consistency with the distance of 140 $\pm$ 8 pc derived
above.

\section{Comparison with Models} \label{sec:models}

We compare in this section the masses and radii of the components of GU
Boo to the predictions by the models of Pietrinferni et al. (2004), Yi et
al. (2001), Girardi et al. (2000), Baraffe et al. (1998), D'Antona \&
Mazzitelli (1997), and Siess et al. (1997). This list of models, although
not complete, provides a representative sample of the predictions by all
current models. The models by Baraffe et al. (1998) and Siess et al.
(1997) are in fact the only ones that explicitly attempt to reproduce the
properties of main sequence low-mass stars. The models by D'Antona \&
Mazzitelli (1997) focus on the pre-main sequence stage, and only include
isochrones up to 0.5 Gyr. The rest of the models, designed for more
massive stars, are not expected to reproduce the observations very well,
but we include them here to illustrate their behavior versus the
observations. For completeness, we also compare the observed mass-log
$T_{\rm eff}$, mass-$M_{V}$, and $\log T_{\rm eff}$-$M_{V}$ relations of
GU Boo to the predictions by the models. We also include in these
comparisons the other known low-mass binaries, whose parameters are listed
in Table \ref{tab:Binaries}.

\begin{figure}[t]
\epsscale{1.0}
\plotone{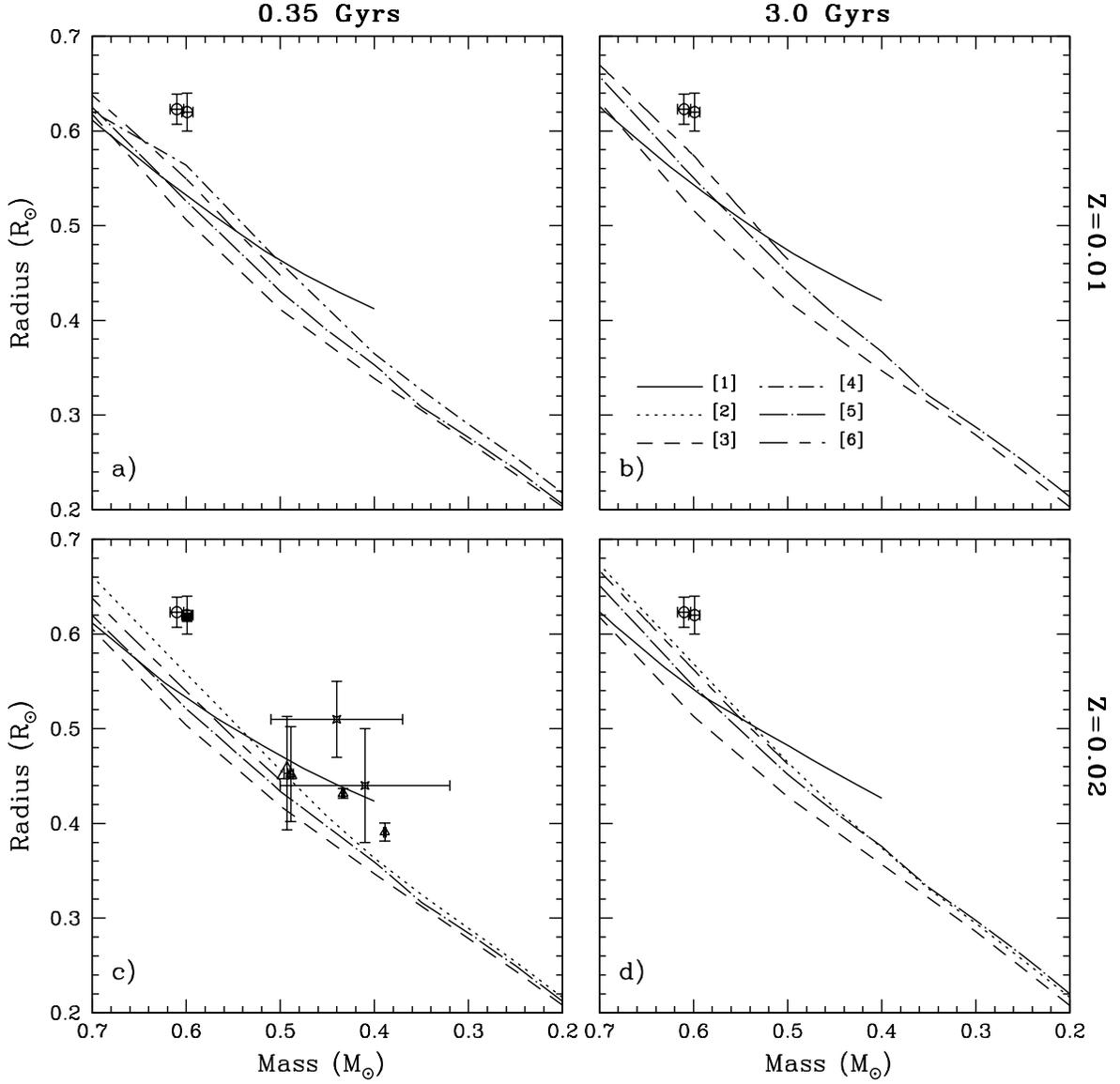}
\caption{Mass-radius relations of stars between 0.7 and 0.2 $M_{\odot}$
predicted by the models in Table \ref{tab:Models}. Each column corresponds to a different
isochrone age (0.35 Gyr on the left, and 3.0 Gyr on the right), while each row
represents a different metallicity (Z = 0.01 at the top, and Z = 0.02 at the
bottom). Each trace corresponds to a different model, with the labels [1],[2],
[3], ... in diagram b) matching the labels in column 1 of Table \ref{tab:Models}. The open
circles show the location of the stars in GU Boo. In diagram c), the filled
circle (overlapping with the less massive component of GU Boo), open triangles,
and crosses mark, respectively, the location of the components of YY Gem, CU
Cnc, and BW3 V38. The components of the recently found TrES-Her0-07621 are shown as open triangles.}
\label{fig:mr} 
\end{figure}

\begin{figure}[t]
\epsscale{1.3}
\plotone{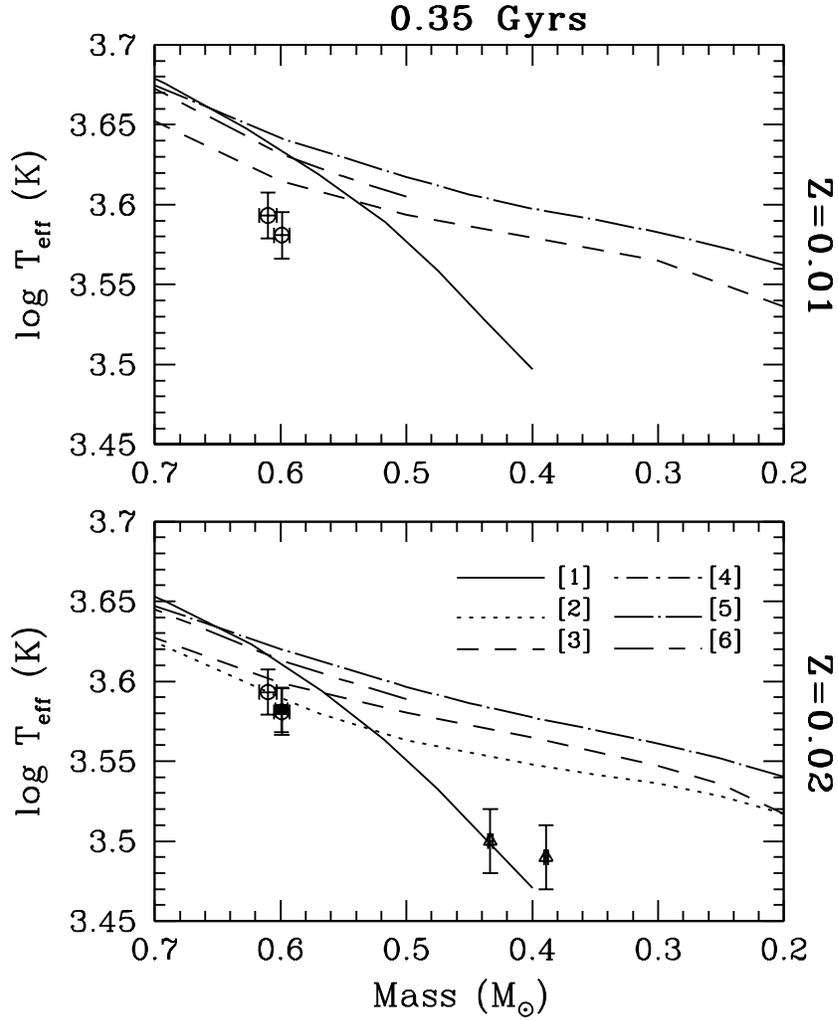}
\caption{Mass-log $T_{\rm eff}$ relations of stars between 0.7 and 0.2 
$M_{\odot}$ predicted by the models in Table \ref{tab:Models}, for an 
isochrone age of 0.35 Gyr. The metallicity in each diagram is Z= 0.01 (top), 
and Z= 0.02 (bottom).
Each trace represents a different model, following the same convention as in
Figure \ref{fig:mr}. The open circles in each diagram represent the location 
of the stars in GU Boo. In the bottom diagram, the filled circle and the open 
triangles mark, respectively, the location of the components of YY Gem and CU 
Cnc.}
\label{fig:mTT} 
\end{figure}

\begin{figure}[t]
\epsscale{1.3}
\plotone{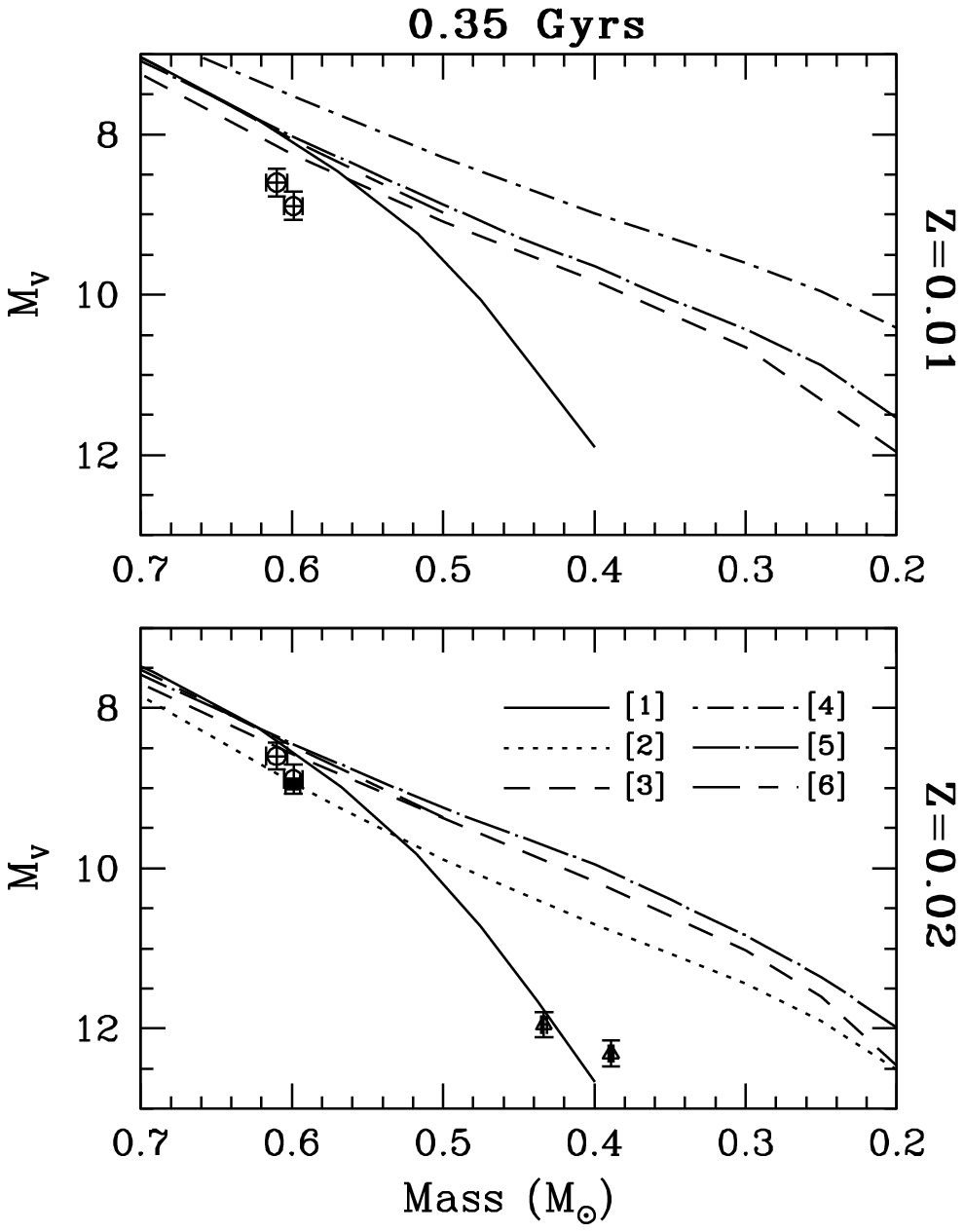}
\caption{Mass-$M_{V}$ relations of stars between 0.7 and 0.2 $M_{\odot}$
predicted by the models in Table \ref{tab:Models}, for an isochrone age of 
0.35 Gyr. See caption to Fig. 8 for a full description of the plots.} 
\label{fig:mM}
\end{figure}

\begin{figure}[t]
\epsscale{1.3}
\plotone{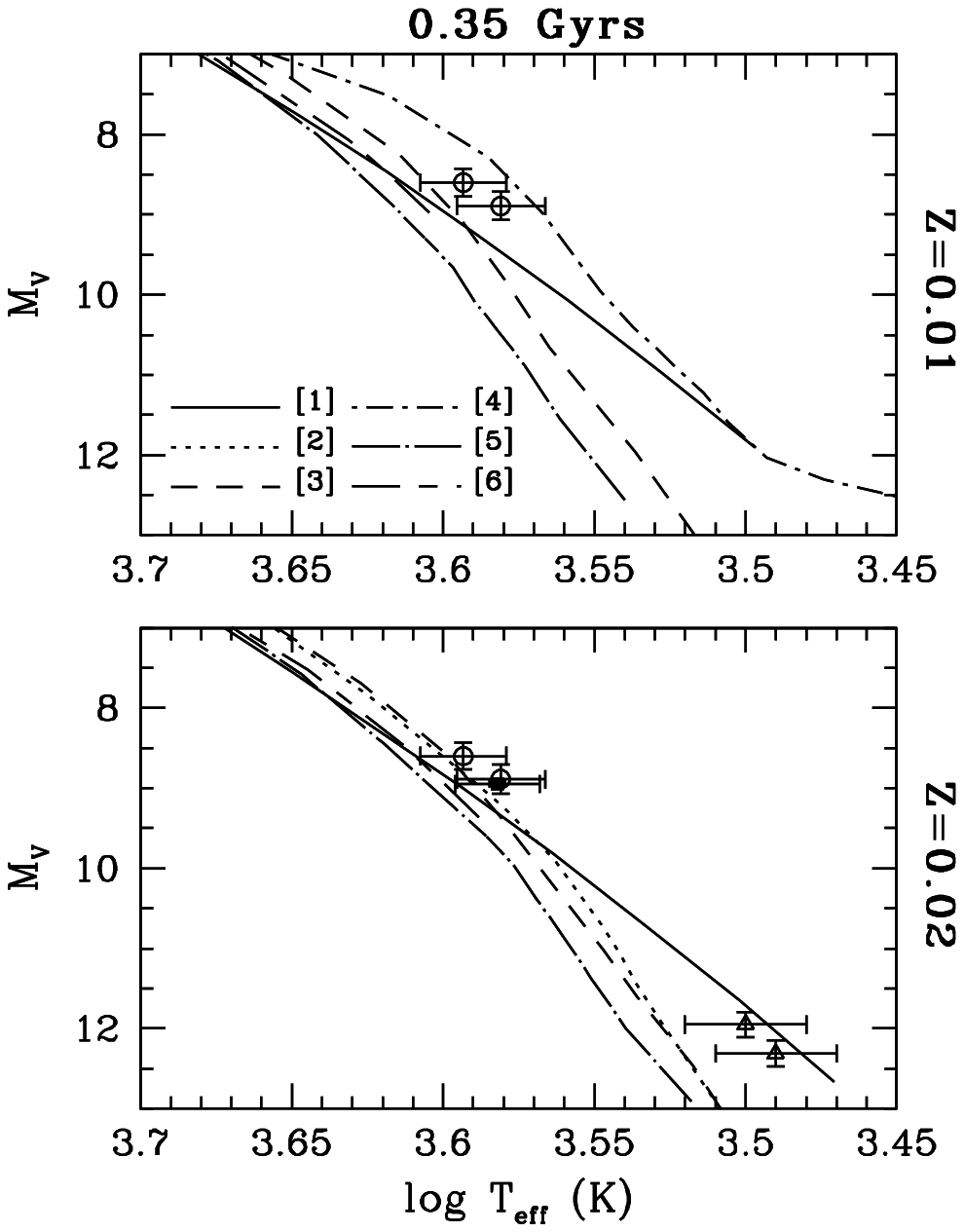}
\caption{$\log T_{\rm eff}$-$M_{V}$ relations of stars between 0.7 and 0.2
$M_{\odot}$ predicted by the models in Table \ref{tab:Models}, for an isochrone
age of 0.35 Gyr. See caption to Fig. 8 for a full description of the plots.}
\label{fig:TM}
\end{figure}

Given that the metal abundance of GU Boo is currently undetermined, we
considered models for two different metallicities, Z= 0.01 and Z= 0.02 (we
expect the metallicity of GU Boo to be close to solar, as explained in \S
\ref{sec:age}). We also considered two main-sequence isochrone ages in the
mass-radius relation (Figure \ref{fig:mr}), after concluding from the
space velocities computed in \S \ref{sec:age} that the system has most
likely already reached the main sequence stage. The age of the isochrones
are 0.35 and 3.0 Gyr, respectively\footnote{Chabrier \& Baraffe (1995)
have estimated a time of arrival to the ZAMS of about 0.3 Gyr for stars of
$\sim$ 0.6 $M_{\odot}$. Therefore both the 0.35 and 3.0 Gyr isochrones
represent stars that have already reached the main sequence stage.}. In
the figures representing the mass-$\log T_{\rm eff}$, mass-$M_{V}$, and
$\log T_{\rm eff}$-$M_{V}$ relations (Figures \ref{fig:mTT}, \ref{fig:mM},
and \ref{fig:TM}) we opted for showing only the 0.35 Gyr isochrones, after
verifying that both the 0.35 and 3.0 Gyr isochrones produce, as expected,
practically identical results. Using the 0.35 Gyr isochrones allows us to
directly compare the parameters of GU Boo to those of YY Gem and CU Cnc,
whose ages have been estimated to be of the order of 320--370 Myr (Torres
\& Ribas 2002; Ribas 2003).

Figure \ref{fig:mr} shows the comparison of the masses and radii of GU Boo
to the different mass-radius relations predicted by the six models
considered in this analysis. Each column in the figure represents a
different age (0.35 Gyr on the left, and 3.0 Gyr on the right), and each
row represents a different metallicity (0.01 at the top, and 0.02 at the
bottom). Details about the parameters of each model are listed in Table
\ref{tab:Models}. The open circles in each plot show the location of the
two components of GU Boo. In diagram c) we also include the location of
the components of YY Gem (filled circle coincident with the less massive
component of GU Boo), CU Cnc (open triangles), and BW3 V38 (crosses). The
age and metallicity in that diagram (Age=0.35 Gyr; Z=0.02) resemble the
age and metallicity of YY Gem and CU Cnc estimated by Torres \& Ribas
(2002) and Ribas (2003), respectively. The stars in BW3 V38 are included
only for completeness, since the current uncertainties of the parameters
of this system are too large to place any strong constraint in the models.
Finally, we have not included CM Dra in that diagram since that binary is
believed to be an older Population II system.

We find that all the models consistently predict smaller radii than the
ones observed, independently of age and metallicity. The discrepancies
between the models and the observations are of the order of 10--15\%. This
result fully agrees with the trend observed in the components of YY Gem
and CU Cnc by Torres \& Ribas (2002) and Ribas (2003).

In Figures \ref{fig:mTT}, \ref{fig:mM}, and \ref{fig:TM} we compare,
respectively, the mass-$\log T_{\rm eff}$, mass-$M_{V}$, and $\log T_{\rm
eff}$-$M_{V}$ relations of GU Boo again to the predictions by the models.
The isochrones in this case are for a single age of 0.35 Gyr (the 3.0 Gyr
isochrones give practically identical results). We consider again the two
values of the metallicity Z= 0.01 and Z= 0.02. The bottom diagrams in each
figure (age= 0.35 Gyr, Z= 0.02) include YY Gem and CU Cnc for comparison

The error bars of the absolute magnitudes and effective temperatures in
these plots are relatively larger than the errors in the masses and radii
in Figure \ref{fig:mr}. This is is consequence of the need to use external
calibrations to compute $T_{\rm eff}$ and $M_{V}$, whereas M and R are
measured directly from the light and radial velocity curves. Because of
the larger errors, the data in the mass-$\log T_{\rm eff}$, mass-$M_{V}$,
and $\log T_{\rm eff}$-$M_{V}$ relations do not provide as strong
constraints on the models as the the mass-radius relation in Figure
\ref{fig:mr}. However, we can still study whether significant
discrepancies between models and observations exist.

The mass-log $T_{\rm eff}$ relations in Figure \ref{fig:mTT} show that such
discrepancies indeed exist between the effective temperatures of GU Boo
and the temperatures predicted by the models, especially in the case of Z
= 0.01.  In the case of Z = 0.02, there is better agreement between models
and observations, although the majority of the models still overestimate
the stellar temperatures by several hundred degrees. Only the models by
Baraffe et al. (1998) give rather close values to the effective
temperatures of GU Boo.  Although this latter model does not reproduce the
temperatures of the individual stars, it does predict a value of $T_{\rm
eff}$= 3890 K for an intermediate mass of 0.6 $M_{\sun}$, which agrees
with the mean effective temperature of GU Boo derived in \S \ref{sec:teff}
($T_{\rm eff}$ = 3870 $\pm$ 130 K). The figure also shows that the
effective temperatures of GU Boo are very similar to the values derived
for YY Gem by Torres \& Ribas (2002).

In the case of the mass-$M_{V}$ relations (Figure \ref{fig:mM}), the Z =
0.02 models are again the ones reproducing more closely the observations,
although none of them provides an exact match to the estimated absolute
magnitudes of GU Boo. There is only a slight difference between the
absolute magnitude of YY Gem ($M_{V}$ = 8.95 $\pm$ 0.03; Torres \& Ribas
2002) and the absolute magnitude of the almost identical secondary in GU
Boo. Such slight difference indicates no large systematic errors between
the two methods used to compute the magnitudes of each binary. The
absolute magnitudes of YY Gem were computed by Torres \& Ribas (2002)
using the Hipparcos parallax of stars in the Castor group (YY Gem is
thought to be part of it), while the absolute magnitudes of GU Boo were
derived using bolometric corrections from models (see \S
\ref{sec:params}). The Baraffe et al. (1998) models reproduce well the
absolute magnitude of YY Gem, but slightly underestimate the magnitudes of
GU Boo.

Finally, the comparison in Figure \ref{fig:TM} of the $\log T_{\rm
eff}$-$M_{V}$ diagram of GU Boo with the models leads to the same
conclusions as Figures \ref{fig:mTT} and \ref{fig:mM}. This is because the
log $T_{\rm eff}$-$M_{V}$ relation is just a combination of the relations
in those two figures. For a given temperature all the Z = 0.02 models
(which produce the best agreement)  overestimate the effective temperature
of the stars, with Baraffe et al. (1998) being the models that yield the
closest match to the observations.

\section{Summary and Conclusions} \label{sec:conclusions}

We report in this paper the finding of GU Boo, a new double-lined,
detached eclipsing binary composed of two M-type stars. We have performed
a detailed analysis of complete R- and I-band light curves and radial
velocity curves of the system to derive its absolute physical stellar
parameters, such as the masses $M_{1}$= 0.610 $\pm$ 0.007 $M_{\sun}$,
$M_{2}$= 0.599 $\pm$ 0.006 $M_{\sun}$, and radii $R_{1}$= 0.623 $\pm$
0.016 $R_{\sun}$, $R_{2}$= 0.620 $\pm$ 0.020 $R_{\sun}$. The uncertainties
of the derived radii of the GU Boo components ($\sim$3\%) are somewhat
larger than expected from the quality of the light curves. This is because
of the complications in the modeling caused by presence of large {\it
spots} on the surface of the stars. We are in the process of collecting
new light curves at near-infrared wavelengths, which are much less
affected by spots, which will allow us to refine the current radius
measurements. The effective temperatures of the components have been
estimated to be ${T_{\rm eff}}_1$= 3920 $\pm$ 130 K and ${T_{\rm eff}}_2$=
3810 $\pm$ 130 K from the observed photometry and by means of both
empirical and model-dependent calibrations. In addition, we have used the
space velocities of the binary to conclude that GU Boo is most likely in
its main sequence evolutionary phase, with an estimated age of several
gigayears. Therefore, the parameters of the stars in GU Boo should be
suitable to test stellar structure models.

The stars in GU Boo are very similar to the components of YY Gem. This
eclipsing binary, together with CU Cnc and CM Dra, were to date the only
simultaneous measurements of stellar masses and radii of M-type main
sequence stars, with uncertainties of less than 3\%. The measurements from
those three binaries were until now the only available data points to
provide critical constraints to low-mass stellar models. The addition of
the two masses and radii from the stars in GU Boo therefore increases by
40\% the number of accurate measurements in that mass regime.

The comparison of the masses and radii of GU Boo with several current
low-mass models of stellar structure corroborates the results of similar
studies of YY Gem and CU Cnc by Torres \& Ribas (2002) and Ribas (2003).
That is, the radii of the stars in this mass domain are at least 10--15\%
larger than the predictions by any of the models. We arrived to the same
result independently of the adopted metallicity of the system. The models
that provide a better fit to the observations are those of Baraffe et al.
(1998), for a metallicity of Z = 0.02, although a noticeable
underestimation of the observed radii is still present. Note that the
metallicity of GU Boo is presently undetermined. This, however, does not
alter our conclusions about the radius discrepancy because of the relative
insensitivity of this parameter to the metallicity. The radii of the stars
are mainly determined by the interior equation of state and depend only
weakly on the atmospheric properties, where the metallicity plays the most
important role.  The mass-radius relation is therefore providing us with
direct information about the interior conditions of low-mass stars.

Following this reasoning, the radius discrepancies between the predictions
of the models and the observations might be, in principle, attributable to
inaccuracies in the equation of state. However, the models successfully
describe the masses and luminosities of the eclipsing binary components
and also of stars in non-eclipsing spectroscopic binaries (e.g., Delfosse
et al. 2000). This apparent paradox can be explained if the stars have
larger radius and cooler temperature than predicted by models but just in
the right proportions. That is, the 10--15\% radius undererestimation
compensated by a 5--7\% temperature overestimation to yield identical
luminosities.

In all the comparisons with models we assume that the eclipsing binary
sample available thus far is representative of the overall M star
population. Because of the presence of discrepancies between models and
theory it is worth reviewing this assumption. The three eclipsing binaries
used in the model comparison are well-detached systems with components
that have not suffered major interactions through mass exchange. There is,
however, one difference with respect to field stars and stars in wide
binaries, which is the relative fast rotation rate caused by orbital
synchronization (periods of 0.5--2.8 days). This high rotational velocity
gives rise to enhanced magnetic activity and thus to the appearance of
surface inhomogeneities, emission lines and X-ray fluxes. All these
phenomena are observed at their peak (saturated activity; Vilhu \& Walter
1987; Stauffer et al. 1994) in YY Gem, CU Cnc and GU Boo. It might be
speculated that the larger radii and lower temperatures could be a reflex
of such enhanced activity. Perhaps the significant spot areal coverage
observed in the three eclipsing systems has the effect of lowering the
overall photospheric temperature, which the star compensates by increasing
its radius to conserve the total radiative flux.

If this scenario turns out to be correct, and there is a correlation
between the radius and the activity level of an M-type star, the
consequences are far reaching. Young low-mass stars are known to be very
active because they are generally fast rotators (e.g., Pizzolato et al.
2003). Then, they lose angular momentum via magnetic braking and spin down
with age. Thus, the radius discrepancy would not only affect the
components of close binaries but any active field M-type star, including
those in young clusters. Only old, inactive M stars (e.g., in the halo)
might have observed and predicted radii in agreement, but no observations
are available for a check.

Clearly, more observational data (in the form of a large sample of
eclipsing binaries with different rotational velocities) and careful
modeling are needed to validate this scenario. This paper summarizes the
first results of an ongoing effort to provide a well-sampled empirical
mass-radius relation of stars below 1 $M_{\sun}$ using detached eclipsing
binaries. Our current sample consists of over a dozen new binaries with
components between $\sim$ 0.35--1.0 $M_{\sun}$, and orbital periods of
less than 2.5 days. The resulting ensemble of stars will provide valuable
tests to low-mass stellar models that should help to resolve the current
lingering discrepancies between observations and theory.

\section{Acknowledgments}

We would like to thank Dr. Scott Shaw from the University of Georgia in
Athens, and the Southeastern Association for Research in Astronomy (SARA)
Observatory for making their allocated telescope time and their facilities
available to us. We give special thanks to Dr. Chris Clemens from the
University of North Carolina at Chapel Hill for his useful advise in the
initial parts of this work. We are also grateful to Drs. Latham and Torres
from CfA-Harvard for their hospitality and for allowing us to used their
implementation of TODCOR. We thank Dr. David Fern\'andez for his comments
and help on performing the kinematic analysis of GU Boo. M.~L-M.
acknowledges research and travel support from the Carnegie Institution of Washington through a
Carnegie Fellowship. I.~R. acknowledges support from the Spanish
Ministerio de Ciencia y Tecnolog\'{\i}a through a Ram\'on y Cajal
Fellowship. An anonymous referee is thanked for a number of useful
comments and suggestions. This publication makes use of data products from
the Two Micron All Sky Survey, which is a joint project of the University
of Massachusetts and the Infrared Processing and Analysis
Center/California Institute of Technology, funded by the National
Aeronautics and Space Administration and the National Science Foundation.
This project has been partially supported by the National Science
Foundation through grant AST 00-94289.

\newpage

\begin{table}[t]
\centering
\footnotesize
\caption{Magnitudes of GU Boo collected from the literature}
\label{tab:Mlit} 
\begin{tabular}{lcc}
\hline\hline
Bandpass&Magnitude&Source\\
\hline
$V_{\rm rotse}$&13.7&Akerlof et al. (2000)\\
I& 11.80$\pm$0.05&This work\tablenotemark{a}\\
b&15.0&USNO-A1.0 Catalog (Monet et al. 1997)\\
r&13.2&USNO-A1.0 Catalog (Monet et al. 1997)\\
B1&15.18&USNO-B1.0 Catalog (Monet et al. 2003)\\
R1&13.12&USNO-B1.0 Catalog (Monet et al. 2003)\\
$J_{\rm 2MASS}$&11.046$\pm$0.024&2MASS All-Sky Catalog of Point Sources (Cutri et al. 2003)\\
$H_{\rm 2MASS}$&10.362$\pm$0.030&2MASS All-Sky Catalog of Point Sources (Cutri et al. 2003)\\
$K_{\rm 2MASS}$&10.222$\pm$0.021&2MASS All-Sky Catalog of Point Sources (Cutri et al. 2003)\\
\hline\hline
\end{tabular}
\tablenotetext{a}{Using the 0.2-m Pisgah Survey robotic telescope (L\'opez-Morales \& Clemens 2004) equipped with an I-band filter.}\\
\end{table}

\begin{table}[t]
\centering
\footnotesize
\caption{Mean effective temperature estimations of GU Boo}
\label{tab:Teff} 
\begin{tabular}{lcc}
\hline\hline
Empirical calibrations&Color Indexes& $T_{\rm eff}$ (K)\\
\hline
Arribas \& Mart\'{\i}nez Roger (1989)&(V-K)&4040\\
Bessell (1979)&(V-I), (R-I), (B-V)&3760\\
\hline\hline
 & & \\
\hline\hline
Model-dependent calibrations&Color Indexes& $T_{\rm eff}$ (K)\\
\hline
Bessel et al. (1998) & (V-I), (V-K), (J-K)& 3835\\
Lejeune et al. (1998) & (V-I), (V-K)& 3760\\
Hauschildt et al. (1999) & (V-I), (V-K), (J-K)& 3900\\
\hline\hline
\end{tabular}
\end{table}

\begin{table}[t]
\centering
\footnotesize
\caption{Parameters of GU Boo derived from the orbital solution
of the radial velocity curves}
\label{tab:Params} 
\begin{tabular}{clc}
\hline\hline
&Element & Value\\
\hline
Adjusted quantities:&&\\
&P (days)\tablenotemark{*}................................. & 0.4887280\\
&$\gamma$ (km s$^{-1}$) ............................ & $-$24.57 $\pm$ 0.36\\
&$K_{1}$(km s$^{-1}$) ........................... & 142.65 $\pm$ 0.66\\
&$K_{2}$(km s$^{-1}$) ........................... & 145.08 $\pm$ 0.73\\
&e\tablenotemark{*} ............................................ & 0.0000\\
&$T_{\circ}$ (HJD) ........................... & 2452723.9811\\
Derived quantities:&&\\
&$M_{1} \sin^3 i$ ($M_{\odot}$) ....................... & 0.6082 $\pm$ 0.0067\\
&$M_{2} \sin^3 i$ ($M_{\odot}$) ....................... & 0.5980 $\pm$ 0.0063\\
&q $\equiv$ $M_2/M_1$ ........................... & 0.9832 $\pm$ 0.0069\\
&$a_{1} \sin i$ ($10^{6}$ km) .................. & 0.9587 $\pm$ 0.0045\\
&$a_{2} \sin i$ ($10^{6}$ km) .................. & 0.9750 $\pm$ 0.0049\\
Other quantities:&&\\
&$N_{obs}$ ....................................... & 103 \\
&Time span (days) ................... & 9.30\\
&$N_{cycles}$ .................................... & 19.0\\
&$\sigma_{1}$ (km s$^{-1}$) ........................... & 4.77\\
&$\sigma_{2}$ (km s$^{-1}$) ........................... & 5.24\\
\hline\hline
\end{tabular}
\tablenotetext{*}{Parameter fixed beforehand}
\end{table}

\begin{table}[t]
\begin{center}
\footnotesize
\caption{Radial velocity measurements}
\label{tab:RVs}
\begin{tabular}{cccccc}
\hline\hline
HJD&Orbital phase&$RV_{1}$&$RV_{2}$&$(O-C)_{1}$&$(O-C)_{2}$\\
&&kms$^{-1}$&km s$^{-1}$&km s$^{-1}$&km s$^{-1}$\\
\hline
2452769.6797&0.505& -28.27& -26.97& -7.86&  1.83\\
2452769.6841&0.514 & -7.43& -36.39&  4.92&  0.60\\
2452769.6867&0.520 &  5.39& -39.76& 13.00&  2.05\\
2452769.6892&0.525 &  3.23& -49.02&  6.30& -2.59\\
2452769.6918&0.530 & 14.17& -49.33& 12.54&  1.88\\
\nodata&\nodata&\nodata&\nodata&\nodata&\nodata\\
\hline\hline
\end{tabular}
\end{center} 
\end{table}

\begin{table}[t]
\centering
\footnotesize
\caption{Times of minima measured from the R- and I-band light curves}
\label{tab:Tmin} 
\begin{tabular}{cccc}
\hline\hline
$E$& Min Type\tablenotemark{*}& HJD &(O-C) (days)\\
\hline
0 &p&2452723.98143&+0.00035\\
2 &p&2452724.95853&$-$0.00001\\
18&p&2452732.77768&$-$0.00051\\
21&s&2452734.00061&+0.00060\\
53&p&2452749.88383&+0.00017\\
80&s&2452762.83524&+0.00028\\
86&s&2452765.76735&+0.00002\\
\hline\hline
\end{tabular}
\tablenotetext{*}{p = primary; s = secondary.}
\end{table}

\begin{table}[t]
\begin{center}
\footnotesize
\caption{Date, phase coverage, and number of observations per phase interval of
the light curve observations}
\label{tab:PhotNights}
\begin{tabular}{cccc}
\hline\hline
Filter&HJD-2450000&Orbital phase&\# of Obs.\\
\hline
R&2733.753--2734.029&0.996--0.559&107\\
R&2749.756--2750.007&0.685--0.253&163\\
R&2765.669--2765.811&0.301--0.901&95\\
\hline
I&2724.873--2725.029&0.825--0.145&190\\
I&2732.767--2733.029&0.977--0.514&178\\
I&2762.688--2762.952&0.201--0.741&254\\
\hline\hline
\end{tabular}
\end{center} 
\end{table}

\begin{table}[t]
\begin{center}
\footnotesize
\caption{R-band light curve measurements}
\label{tab:LCsR}
\begin{tabular}{ccc}
\hline\hline
HJD&Orbital phase&Differential mag.\\
\hline
2452733.75358&0.996&0.222\\
2452733.75571&0.000&0.152\\
2452733.75903&0.007&0.206\\
2452733.76210&0.014&0.296\\
2452733.76833&0.026&0.498\\
2452733.77785&0.046&0.735\\
2452733.78065&0.051&0.783\\
2452733.78310&0.056&0.827\\
\nodata&\nodata&\nodata\\
\hline\hline
\end{tabular}
\end{center} 
\end{table}

\begin{table}[t]
\centering
\scriptsize
\caption{Parameters of GU Boo for two spot scenarios}
\label{tab:PhotSols}
\begin{tabular}{llccc}
\hline\hline
&Parameter& Spot Scenario \# 1&Spot Scenario \# 2&Fixing\tablenotemark{\dagger}\\
\hline
Geometric& P (days)\tablenotemark{*}  ......................&0.4887280&0.4887280&[1]\\
parameters:& e\tablenotemark{*}  ..................................&0.0000&0.0000&[1]\\
& $\Delta\phi$  ................................&0.0012 $\pm$ 0.0002 &0.00007 $\pm$ 0.00002&\\
& i (deg)  ...........................& 88.2 $\pm$ 0.2 & 87.6 $\pm$ 0.2&\\
& $\Omega_{1}$ ...............................& 5.523 $\pm$ 0.118 & 5.382 $\pm$ 0.112 &\\
& $\Omega_{2}$ ...............................& 5.636 $\pm$ 0.112 & 5.427 $\pm$ 0.105&\\
& q = $M_{2}/M_{1}$\tablenotemark{*}  ................ & 0.9832 & 0.9832 &[1]\\
Fractional radii& ${r_{1}}_{\rm point}$  ..........................& 0.2256 $\pm$ 0.0016 & 0.2292 $\pm$ 0.0016&\\
of primary& ${r_{1}}_{\rm pole}$  .............................& 0.2188 $\pm$ 0.0014 & 0.2220 $\pm$ 0.0018&\\
& ${r_{1}}_{\rm side}$  .............................& 0.2211 $\pm$ 0.0015 & 0.2244 $\pm$ 0.0015&\\
& ${r_{1}}_{\rm back}$  ............................& 0.2243 $\pm$ 0.0016 &  0.2279 $\pm$ 0.0016&\\
& $r_{1}$\tablenotemark{**} ..................................& 0.221 $\pm$ 0.005 & 0.224 $\pm$ 0.005&\\
Fractional radii& ${r_{2}}_{\rm point}$  ............................& 0.2366 $\pm$ 0.0014& 0.2282 $\pm$ 0.0021&\\
of secondary& ${r_{2}}_{\rm pole}$  .............................& 0.2281 $\pm$ 0.0012 & 0.2209 $\pm$ 0.0018&\\
& ${r_{2}}_{\rm side}$  .............................& 0.2309 $\pm$ 0.0013 & 0.2234 $\pm$ 0.0019&\\
& ${r_{2}}_{\rm back}$  .............................& 0.2349 $\pm$ 0.0014 & 0.2268 $\pm$ 0.0020&\\
& $r_{2}$\tablenotemark{**} ...................................& 0.231 $\pm$ 0.004 & 0.223 $\pm$ 0.006&\\
Radiative& Gravity brightening\tablenotemark{*}  .........& 0.2 & 0.2 &[2]\\
parameters:& ${T_{\rm ph}}_1$ (K)\tablenotemark{*} ....................& 4050 & 3940&[4] \\
& ${T_{\rm ph}}_2$ (K) ....................& 3820 $\pm$ 10 & 3800 $\pm$ 10&\\
& Albedo\tablenotemark{*}  ............................& 0.5 & 0.5&[2]\\
Light ratio (R-band)& $L_{2}/L_{1}$\tablenotemark{***}  ..........................& 0.90 $\pm$ 0.04 & 0.80 $\pm$ 0.04&\\
Light ratio (I-band)& $L_{2}/L_{1}$\tablenotemark{***}  ..........................& 0.96 $\pm$ 0.03 & 0.85 $\pm$ 0.03 &\\
Limb darkening & & & &\\
 coefficients \tablenotemark{*} :& ${x_{\rm bol}}_1$, ${y_{\rm bol}}_1$  ........................ & 0.207, 0.665 & 0.210, 0.662&[2]\\
(square-root law):& ${x_{\rm bol}}_2$, ${y_{\rm bol}}_2$  ........................ & 0.192, 0.719 & 0.201, 0.695&[2]\\
& (${x_{\rm R}}_{1}$, ${y_{\rm R}}_{1}$)      ......................... & 0.607, 0.212 & 0.617, 0.202&\\
& (${x_{\rm R}}_{2}$, ${y_{\rm R}}_{2}$)      ......................... & 0.463, 0.369 & 0.510, 0.316&\\
& (${x_{\rm I}}_{1}$, ${y_{\rm I}}_{1}$)      .......................... & 0.193, 0.618 & 0.201, 0.610 &\\
& (${x_{\rm I}}_{2}$, ${y_{\rm I}}_{2}$)      .......................... & 0.059, 0.784 & 0.095, 0.736 &\\
Spot 1& Star Location .. .............& Primary & Primary&\\
parameters:& Latitude (deg)\tablenotemark{*}  ..............& 45 & 63 &[2],[3]\\
& Longitude (deg)  .............& 349 & 349 &\\
& Angular radius (deg)  .....& 89 & 33&\\
& $T_{\rm spot}/T_{\rm surf}$\tablenotemark{*}  ...................& 0.95 & 0.94&[4]\\
Spot 2& Star Location  .............& Primary & Secondary&\\
parameters:& Latitude (deg)\tablenotemark{*}  ..............& 45 & 63 &[2],[3]\\
& Longitude (deg)  .............& 153 & 22 &\\
& Angular radius (deg)  ......& 43 & 23&\\
& $T_{\rm spot}/T_{\rm surf}$\tablenotemark{*}  ...................& 0.94 & 1.06 &[4]\\
&&&&\\
rms residuals &R-band&0.0087& 0.0086&\\
              &I-band&0.0115& 0.0118&\\
\hline\hline
\end{tabular}
\tablenotetext{\dagger}{[1]: Parameter fixed by the orbital solution of the radial velocity curve; [2]: Parameter fixed from the literature; [3]: Parameter fixed from preliminary solutions of WD2003; [4]: Other sources (see text).}
\tablenotetext{*}{Parameters fixed in the final computation of the orbital solution (see explanation in the text).}
\tablenotetext{**}{Volume radius.}
\tablenotetext{***}{Out-of-eclipse average (phase intervals: 0.075-0.427 \& 0.574-0.927.}
\end{table}

\begin{table}[t]
\centering
\footnotesize
\caption{Absolute dimensions and main physical parameters of the
components of GU Boo}
\label{tab:AbsDim} 
\begin{tabular}{lcc}
\hline\hline
Parameter & Primary & Secondary\\
\hline
Mass ($M_{\odot}$)  ................ & 0.610 $\pm$ 0.007& 0.599 $\pm$ 0.006\\
Radius ($R_{\odot}$)  .............. & 0.623 $\pm$ 0.016& 0.620 $\pm$ 0.020\\
$\log g$ (cgs)  .................. & 4.634 $\pm$ 0.023& 4.630 $\pm$ 0.028\\
$v_{\rm sync}$ sini (km s$^{-1}$)  ....& 64.43 $\pm$ 1.65 & 64.15 $\pm$ 2.07\\
$T_{\rm eff}$ (K)  .................... & 3920 $\pm$ 130& 3810 $\pm$ 130\\
$L/L_{\odot}$  ......................... & 0.082 $\pm$ 0.011& 0.073 $\pm$ 0.013\\
$M_{\rm bol}$ (mag)  ................ & 7.46 $\pm$ 0.15& 7.60 $\pm$ 0.16\\
$M_{V}$ (mag) . ................. & 8.60 $\pm$ 0.17& 8.89 $\pm$ 0.18\\
\hline\hline
\end{tabular}

NOTES -- The luminosities and bolometric magnitudes were computed from the
radii and the effective temperatures, using $M_{\rm bol}^{\odot}$ = 4.72 and
$T_{\rm eff}^{\odot}$= 5778 K.  The absolute magnitudes $M_{V}$ were computed
using bolometric corrections derived from the models in Table \ref{tab:Models}.
\end{table}

\begin{table}[t]
\centering
\footnotesize
\caption{Summary of the main physical parameters of the six M-type
eclipsing binaries studied to date.}
\label{tab:Binaries} 
\begin{tabular}{llccc}
\hline\hline
Binary&Parameter& Primary & Secondary & Reference\\
\hline
YY Gem  .....&Mass ($M_{\odot}$)  .......&0.5992 $\pm$ 0.0047 & 0.5992 $\pm$ 0.0047&[a]\\
&Radius ($R_{\odot}$)  .....&0.6191 $\pm$ 0.0057 & 0.6191 $\pm$ 0.0057&[a]\\
&$T_{\rm eff}$ (K)  ...........&3820 $\pm$ 100 & 3820$\pm$ 100& [a]\\
&$M_{V}$  ...................&8.950$\pm$ 0.029 & 8.950$\pm$ 0.029& [a]\\
BW3 V38  ....&Mass ($M_{\odot}$)  .......&0.44 $\pm$ 0.07& 0.41 $\pm$ 0.09& [b]\\
&Radius ($R_{\odot}$)  .....&0.51 $\pm$ 0.04 & 0.44 $\pm$ 0.06& [b]\\
&$T_{\rm eff}$ (K)  ...........&$3500^{*}$ & 3448 $\pm$ 11& [b]\\
TrES-Her0  ...&Mass ($M_{\odot}$)  .......&0.493 $\pm$ 0.003& 0.489 $\pm$ 0.003& [c]\\
&Radius ($R_{\odot}$)  .....&0.453 $\pm$ 0.060 & 0.452 $\pm$ 0.050& [c]\\
CU Cnc  ......&Mass ($M_{\odot}$)  .......&0.4333 $\pm$ 0.0017& 0.3890 $\pm$ 0.0014& [d]\\
&Radius ($R_{\odot}$)  .....&0.4317 $\pm$ 0.0052 & 0.3908 $\pm$ 0.0094& [d]\\
&$T_{\rm eff}$ (K)  ...........&3160 $\pm$ 150 & 3125 $\pm$ 150& [d]\\
&$M_{V}$  ...................&11.95 $\pm$ 0.16 & 12.31 $\pm$ 0.16& [d]\\
CM Dra  ......&Mass ($M_{\odot}$)  .......&0.2307 $\pm$ 0.0010& 0.2136 $\pm$ 0.0010& [e]\\
&Radius ($R_{\odot}$)  .....&0.2516 $\pm$ 0.0020 & 0.2347 $\pm$ 0.0019& [e]\\
\hline
GU Boo  ......&Mass ($M_{\odot}$)  .......&0.610 $\pm$ 0.007& 0.599 $\pm$ 0.006& [f]\\
&Radius ($R_{\odot}$)  .....&0.623 $\pm$ 0.016 & 0.620 $\pm$ 0.020& [f]\\
&$T_{\rm eff}$ (K)  ...........&3920 $\pm$ 130 & 3810 $\pm$ 130& [f]\\
&$M_{V}$  ...................&8.60 $\pm$ 0.17 & 8.89 $\pm$ 0.18& [f]\\
\hline\hline
\end{tabular}

NOTES -- [a] Torres \& Ribas (2002), [b] Maceroni \& Montalb\'an (2004), [c] Creevey et al. (2005), 
[d] Ribas (2003), and [e] Metcalfe et al. (1996). The last entries correspond to
parameters of the newest binary, GU Boo, analyzed in this work (reference [f]).
\end{table}

\begin{table}[t]
\centering
\footnotesize
\caption{Summary of the parameters of the models used in the M--R,
M--log$T_{eff}$, M--$M_{V}$, and log$T_{eff}$--$M_{V}$ relations}
\label{tab:Models} 
\begin{tabular}{clccccccl}
\hline\hline
Model&Reference&$M_{\rm min}$&$M_{\rm max}$&Y&Age&$Z_{1}$&$Z_{2}$& Notes\\
ID&&$[M_{\sun}]$&$[M_{\sun}]$&&[Gyr]&(=0.01)&(=0.02)&(*)\\
\hline\hline
 [1] &Yi et al.&0.4&1.0&*&0.35&$\surd$&$\surd$&$Y_{Z_{2}}$=0.250\\
 &(2001)&&&*&3.00&$\surd$&$\surd$&$Y_{Z_{3}}$=0.270\\
\hline
$[2]$ &Baraffe et &0.075&1.0&0.275&0.35&---&$\surd$&\nodata\\
&al. (1998)&&&0.275&3.00&---&$\surd$&\nodata\\
\hline
$[3]$&Siess et al&0.1&1.0&*&0.35&$\surd$&$\surd$&$Y_{Z_{2}}$=0.256\\
&(1997)&&&*&3.00&$\surd$&$\surd$&$Y_{Z_{3}}$=0.277\\
\hline
$[4]$&D'Antona &0.02&1.0&*&0.35&$\surd$&$\surd$&$Y_{Z_{2}}$=0.260\\
& et al. (1997)&&&*&3.00&---&---&$Y_{Z_{3}}$=0.280\\
\hline
$[5]$&Girardi et&0.15&1.0&*&0.35&$\surd$&$\surd$&$Z_{2}$=0.008;$Y_{Z_{2}}$=0.250\\
&al. (2000)&&&*&3.00&$\surd$&$\surd$&$Z_{3}$=0.019;$Y_{Z_{3}}$=0.273\\
\hline
$[6]$&Pietrinferni&0.5&1.0&*&0.35&$\surd$&$\surd$&$Y_{Z_{2}}$=0.259\\
&et. al (2004)&&&*&3.00&$\surd$&$\surd$&$Z_{3}$=0.0198;$Y_{Z_{3}}$=0.273\\
\hline\hline
\end{tabular}
NOTE -- $M_{\rm min}$ and $M_{\rm max}$ correspond to the minimum and maximum
stellar mass simulated by the models in the figure, Y is the He abundance, and
the check marks (or crossed lines) under the values of the metallicities
$Z_{1}$ and $Z_{2}$ indicate whether or not isochrones for that particular
metallicity are available.
\end{table}

\newpage

\clearpage

\end{document}